\documentclass[twocolumn]{aastex631}

\received{April 6, 2023}
\revised{May 26, 2023}
\accepted{June 13, 2023}

\begin{document}

\title{Multiple Stellar Populations in Metal-Poor Globular Clusters with JWST:  a NIRCam view of M\,92}

\correspondingauthor{Tuila Ziliotto}
\email{tuila.ziliotto@unipd.it}

\author{Tuila Ziliotto}
\affiliation{Dipartimento di Fisica e Astronomia Galileo Galilei, Univ. di Padova, Vicolo dell'Osservatorio 3, Padova, IT-35122}

\author{Antonino Milone}
\affiliation{Dipartimento di Fisica e Astronomia Galileo Galilei, Univ. di Padova, Vicolo dell'Osservatorio 3, Padova, IT-35122}
\affiliation{Istituto Nazionale di Astrofisica - Osservatorio Astronomico di Padova, Vicolo dell'Osservatorio 5, Padova, IT-35122}


\author{Anna F. Marino}
\affiliation{Istituto Nazionale di Astrofisica - Osservatorio Astronomico di Padova, Vicolo dell'Osservatorio 5, Padova, IT-35122}
\affiliation{Istituto Nazionale di Astrofisica - Osservatorio Astrofisico di Arcetri, Largo Enrico Fermi, 5, Firenze, IT-50125}

\author{Aaron L. Dotter}
\affiliation{Department of Physics and Astronomy, Dartmouth College, 6127 Wilder Laboratory, Hanover, NH 03755, USA}

\author{Alvio Renzini}
\affiliation{Istituto Nazionale di Astrofisica - Osservatorio Astronomico di Padova, Vicolo dell'Osservatorio 5, Padova, IT-35122}

\author{Enrico Vesperini}
\affiliation{Department of Astronomy, Indiana University, Bloomington, IN 47401, USA}

\author{Amanda Karakas}
\affiliation{School of Physics and Astronomy, Monash University, VIC 3800, Australia}
\affiliation{Centre of Excellence for Astrophysics in Three Dimensions (ASTRO-3D), Melbourne, Victoria, Australia}

\author{Giacomo Cordoni}
\affiliation{Dipartimento di Fisica e Astronomia Galileo Galilei, Univ. di Padova, Vicolo dell'Osservatorio 3, Padova, IT-35122}
\affiliation{Istituto Nazionale di Astrofisica - Osservatorio Astronomico di Padova, Vicolo dell'Osservatorio 5, Padova, IT-35122}

\author{Emanuele Dondoglio}
\affiliation{Dipartimento di Fisica e Astronomia Galileo Galilei, Univ. di Padova, Vicolo dell'Osservatorio 3, Padova, IT-35122}

\author{Maria V. Legnardi}
\affiliation{Dipartimento di Fisica e Astronomia Galileo Galilei, Univ. di Padova, Vicolo dell'Osservatorio 3, Padova, IT-35122}

\author{Edoardo P. Lagioia}
\affiliation{Dipartimento di Fisica e Astronomia Galileo Galilei, Univ. di Padova, Vicolo dell'Osservatorio 3, Padova, IT-35122}

\author{Anjana Mohandasan}
\affiliation{Dipartimento di Fisica e Astronomia Galileo Galilei, Univ. di Padova, Vicolo dell'Osservatorio 3, Padova, IT-35122}

\author{Sarah Baimukhametova}
\affiliation{Dipartimento di Fisica e Astronomia Galileo Galilei, Univ. di Padova, Vicolo dell'Osservatorio 3, Padova, IT-35122}



\begin{abstract}

Recent work on metal-intermediate globular clusters (GCs) with [Fe/H]=$-1.5$ and $-0.75$ has illustrated the theoretical behavior of multiple populations in photometric diagrams obtained with the James Webb Space Telescope (JWST). These results are confirmed by observations of multiple populations among M-dwarfs of 47 Tucanae. Here, we explore the multiple populations in metal-poor GCs with [Fe/H]=$-$2.3.  We take advantage of synthetic spectra and isochrones that account for the chemical composition of multiple populations to identify photometric diagrams that separate the distinct stellar populations of GCs. We derived high-precision photometry and proper motion for main-sequence stars in the metal-poor GC M\,92 from JWST and Hubble Space Telescope ({\it HST}) images. We identified a first generation (1G) and two main groups of second-generation stars (2G$_{\rm A}$ and 2G$_{\rm B}$) and investigated their kinematics and chemical composition. We find isotropic motions with no differences among the distinct populations. The comparison between the observed colors of M\,92 stars and the colors derived by synthetic spectra reveals that helium abundance of 2G$_{\rm A}$ and 2G$_{\rm B}$ stars are higher than that of the 1G by $\Delta Y \sim 0.01$ and $0.04$, respectively.  The $m_{\rm F090W}$ vs.\, $m_{\rm F090W}-m_{\rm F277W}$ color-magnitude diagram shows that below the knee, MS stars exhibit a wide color broadening due to multiple populations.  We constrain the amount of oxygen variation needed to reproduce the observed MS width, which is consistent with results on red-giant branch stars. We conclude that multiple populations with masses of $\sim$0.1-0.8$M_{\odot}$ share similar chemical compositions.

\end{abstract}



\section{Introduction} \label{sec:intro}

In the past few decades, the {\it Hubble Space Telescope} ({\it HST}) has been instrumental in demonstrating that the color-magnitude diagrams (CMDs) of most globular clusters (GCs) consist of multiple sequences of stars.
 Each stellar sequence corresponds to a distinct stellar population with a characteristic chemical composition. GCs typically host distinct groups of stars, one with the same chemical composition as halo field stars (1G), and a second-generation of stars (2G), which are typically enhanced in He, N, Na, and Al and depleted in C and O \citep[see][for recent reviews]{bastian2018a, gratton2019a, marino2019a, milone2022a}.
 
The origin of multiple populations is one of the most debated  topics of modern stellar astrophysics.
Most scenarios of the formation of multiple populations suggest that GCs experienced multiple star formation episodes where 2G stars formed from material polluted by more-massive  1G stars. The nature of the polluters is controversial. The main  candidates include intermediate-mass asymptotic-giant-branch (AGB) and super AGB stars \citep[][]{dantona1983a, dantona2016a, calura2019a,cottrell1981a}, fast-rotating massive stars \citep{decressin2007a}, massive interactive binaries \citep{renzini2022a}, and super-massive stars \citep{denissenkov2014a}. 
 Alternatively, the chemical enrichment of 2G stars could originate from processed gas ejected by massive binary systems or super-massive stars and accreted by pre-MS stars in the proto-GCs \citep{bastian2013a, gieles2018a}.

 Most photometric studies on multiple stellar populations are based on ultraviolet photometry either from {\it HST} \citep[e.g.\,][and references therein]{milone2012a, milone2017a, piotto2015a, lagioia2021a, milone2022a} or from ground-based facilities \citep[e.g.][]{marino2008a, yong2008a, monelli2013, lee2022a, jang2022a}. 
  Indeed, UV filters enclose spectral regions that are affected by molecules that include C, N, and O.
  Specifically, the {\it HST} filter F336W encompasses NH molecular bands, while the F275W filter includes the OH bands. The F410M and F438W filters include CH molecular bands, whereas the
narrow-band F280N filter encloses Mg lines. A similar conclusion can be extended to the Johnson-Cousin and Stroemgren filters \citep{marino2008a, sbordone2011a, milone2020b, vandenberg2022a}. 
 The main limitation of studies based on UV photometry is that they are limited to stars of the upper main sequence (MS) or brighter. Indeed, it is challenging to obtain high-precision UV photometry of faint stars with present-day facilities.
 On the contrary, \textit{HST} photometry in appropriate NIR bands, such as the F110W and F160W filters of the IR channel of the Wide-Field Camera 3 (WFC3), is an efficient tool to identify stellar populations among M-dwarfs \citep{milone2012a, milone2019a, dondoglio2022a}. 

Recently, \citet{milone2023a} computed isochrones of different stellar populations in the JWST/NIRCam filters for metal-rich ([Fe/H]=$-$0.75) and metal-intermediate GCs ([Fe/H]=$-$1.5). The results allowed them to identify photometric diagrams suitable for identifying and characterizing multiple stellar populations.
In particular, NIRCam images are formidable tools to disentangle multiple populations among M-dwarfs, in the CMD region that ranges from the MS knee towards the H-burning limit.
NIRCam observations confirm these theoretical results: \citet{milone2023a} show that the M-dwarfs of 47\,Tucanae span a wide F115W$-$F322W2 color interval.
The color broadening is due to the different amounts of the blanketing from molecules composed of oxygen (mostly water vapor) in the F322W2 filter, as 2G stars have lower oxygen, hence low H$_{2}$O content, compared to 1G stars that are oxygen-, hence H$_{2}$O-rich.
Based on a pseudo-two-color diagram or `Chromosome Map' (ChM), \citet{milone2023a} unveiled an extended first population and three main groups of second-population stars among the M-dwarfs of 47\,Tucanae, with different oxygen abundances. 
Isochrones indicate that the diagrams composed of NIRCam photometry alone are poorly sensitive to multiple populations among stars that are brighter than the MS knee, with the exception of stars close to the tip of the red-giant branch (RGB), where we expect that 1G and 2G stars exhibit different fluxes in certain NIRCam filters, like F277W and F460M. The magnitude difference is primarily due to the effect of H$_2$O and CO molecules in the atmospheres of these stars. 2G stars, which are O-depleted with respect to the 1G, exhibit weaker lines of these molecules, hence brighter fluxes when compared to 1G stars. \citep{salaris2019a, milone2023a}.

In this work, we explore multiple populations with {\it HST} and {\it JWST} at low metallicities ([Fe/H]=$-$2.3). We  first investigate the metal-poor GC M\,92 (NGC 6341; [Fe/H]=$-$2.31, 2010 version of the \citet{harris1996a} catalog) and its multiple stellar populations by using images collected with {\it HST} and {\it JWST}. We compute isochrones that account for the chemical composition of 1G and 2G stars in metal-poor GCs to construct the photometric diagrams based on {\it HST} and NIRCam photometry that allow us to identify and characterize multiple populations at low metallicities.
 The paper is structured as follows. Section\,\ref{sec:data} describes the dataset and the methods to derive high-precision photometry, astrometry, and proper motions. Section\,\ref{sec:MPs} is dedicated to the photometry of multiple stellar populations along the RGB and the MS. Synthetic spectra and isochrones of 1G and 2G stars in metal-poor GCs are presented in Section\,\ref{sec:teo}, where we also infer the chemical composition of the multiple populations of M\,92. Section\,\ref{sec:conclusions} provides the summary of the results and discussion.


\begin{table*}
  \caption{Description of the images used in the paper. For each dataset, we indicate the mission ({\it JWST} or {\it HST}), the camera, the filter, the date, the exposure times, the GO program, and the principal investigator.}
\tabcolsep=0.13cm
\footnotesize
\centering
\begin{tabular}{c c c c c l l}
\hline \hline
 MISSION & CAMERA & FILTER  & DATE & N$\times$EXPTIME & PROGRAM & PI \\
\hline
  HST    &  WFC/ACS &  F814W   &  August, 27, 2002  &  0.5s$+$6s$+$100s      &  9453     &  T.\,M.\,Brown   \\
  HST &  WFC/ACS  & F625W & August, 07, 2004  & 10s$+$3$\times$120s & 10120 & S.\,Anderson \\
  HST &  WFC/ACS  & F658N & August, 07, 2004  & 2$\times$350s$+$2$\times$555s & 10120 &  S.\,Anderson \\
  HST    &  WFC/ACS &  F606W   &  November, 14, 2006  &  7s$+$4$\times$140s    &  10775    &  A.\,Sarajedini   \\  
  HST    &  WFC/ACS &  F814W   &  November, 14, 2006  &  7s$+$4$\times$150s    &  10775    &  A.\,Sarajedini   \\  
HST &  UVIS/WFC3 & F390W & October, 10, 2009 & 2$\times$2s$+$2$\times$348s$+$2$\times$795s & 11664 &   T.\,M.\,Brown\\
HST &  UVIS/WFC3 & F555W & October, 10, 2009 & 1s$+$30s$+$2$\times$665s & 11664 &   T.\,M.\,Brown \\
HST &  UVIS/WFC3 & F336W & October, 11, 2009 & 30s$+$2$\times$425s & 11729 & J.\,Holtzman \\
HST &  UVIS/WFC3 & F390M & October, 11, 2009 & 50s$+$2$\times$700s &  11729 & J.\,Holtzman\\
HST &  UVIS/WFC3 & F390W & October, 11, 2009 & 10s  & 11729 & J.\,Holtzman \\
HST &  UVIS/WFC3 & F395N & October, 10–11, 2009 & 90s$+$2$\times$965s & 11729 & J.\,Holtzman \\
HST &  UVIS/WFC3 & F410M & October, 11, 2009 & 40s$+$2$\times$765s & 11729 & J.\,Holtzman \\
HST &  UVIS/WFC3 & F467M & October, 11, 2009 & 40s$+$2$\times$350s & 11729 & J.\,Holtzman \\
HST &  UVIS/WFC3 & F547M & October, 11, 2009 & 5s$+$40s$+$400s & 11729 & J.\,Holtzman \\
HST &  WFC/ACS  & F475W & August, 21, 2012  & 4$\times$400s & 12116 & J.\,Holtzman \\
HST &  UVIS/WFC3 & F275W & October, 22, 2013 & 2$\times$707s &  13297 & G.\,Piotto\\
HST &  UVIS/WFC3 & F336W & October, 22, 2013 & 2$\times$304s &  13297 & G.\,Piotto\\
HST &  UVIS/WFC3 & F438W & October, 22, 2013 & 59s &  13297 & G.\,Piotto\\
HST &  UVIS/WFC3 & F275W & August, 03, 2014 & 2$\times$819s &  13297 & G.\,Piotto\\
HST &  UVIS/WFC3 & F336W & August, 03, 2014 & 2$\times$304s &  13297 & G.\,Piotto\\
HST &  UVIS/WFC3 & F438W & August, 03, 2014 & 57s &  13297 & G.\,Piotto\\
HST &  UVIS/WFC3 & F225W & June, 28, 2019 & 3$\times$30s$+$2$\times$800s$+$3$\times$805s$+$815s$+$820s$+$4$\times$835s &  15173 & J. S. Kalirai \\
HST &  WFC/ACS &  F814W   &  January, 24, 2021 &  35s$+$4$\times$337s &  16289      &  M.\,Libralato   \\
  JWST   &  NIRCam  &  F090W   &     June, 20-21, 2022  & 4$\times$311s   &  1334        &  D.\,R.\,Weisz  \\
  JWST   &  NIRCam  &  F150W   &     June, 20-21, 2022  & 4$\times$311s   &  1334        &  D.\,R.\,Weisz   \\
  JWST   &  NIRCam  &  F277W   &     June, 20-21, 2022  & 4$\times$311s   &  1334        &  D.\,R.\,Weisz   \\
  JWST   &  NIRCam  &  F444W   &     June, 20-21, 2022  & 4$\times$311s   &  1334        &  D.\,R.\,Weisz   \\
     \hline\hline
\end{tabular}
  \label{tab:data}
 \end{table*}

\section{Data and data reduction}\label{sec:data}
To investigate multiple stellar populations in M\,92, we used both NIRCam data and images collected with the wide-field channel of the Advanced Camera for Surveys (WFC/ACS) and the ultraviolet and visual channel of the Wide Field Camera 3 (UVIS/WFC3) on board {\it HST}. The main properties of the dataset are summarized in Table\,\ref{tab:data}.

We derived the photometry and astrometry of stars in all {\it HST} images with the computer program img2xym, originally developed by \citet{anderson2006a} to reduce WFC/ACS images. In a nutshell, separately, we measured the stellar fluxes and positions in each image by using a spatially
variable point-spread-function (PSF) model plus a ‘perturbation PSF’ that fine-tunes the fitting to account for
small variations of the {\it HST} focus. The latter is derived using unsaturated, bright, and isolated stars, while the magnitudes of saturated stars are calculated as in \citet{gilliland2004a}. 
The various measurements of magnitudes and positions are then averaged together to get the best estimates.
 We used a similar method to calculate the magnitudes and positions of stars in the NIRCam images. The difference is that we derived a spatially-variable PSF model for each image based on the available unsaturated, bright, and isolated stars. To do this, we used the computer program img2psf, originally developed by \citet{anderson2006b} for images collected with the Wide Field Imager of the 2.2m Telescope in La Silla and adapted by \citet{milone2023a} for NIRCam \citep[see][for details]{milone2023a}.

We corrected the stellar positions for the effects of the geometric distortions of the WFC/ACS, UVIS/WFC3, and the Short-Wavelength modules of the NIRCam detectors by adopting the solutions by \citet{anderson2006a}, \citet{bellini2009a, bellini2011a}, and \citet{milone2023a}, respectively.
 We calibrated the photometry into the Vega-mag system as in \citet{milone2022b} and using the zero points provided by the Space Telescope Science Institute webpage\footnote{https://www.stsci.edu/hst/instrumentation/acs/data-analysis/zeropoints; \newline https://www.stsci.edu/hst/instrumentation/wfc3/data-analysis/photometric-calibration; \newline https://jwst-docs.stsci.edu/jwst-near-infrared-camera/nircam-performance/nircam-absolute-flux-calibration-and-zeropoints}.
Since we are interested in stars with high-precision photometry, we followed the recipe by \citet{milone2022b} to select the relatively-isolated stars that are well-fitted by the PSF model and have small photometric and astrometric uncertainties.


\subsection{Proper motions}
 We used multi-epoch {\it HST} and {\it JWST} images and Gaia DR3 data \citep{gaia2021a} to derive the proper motions of the stars in the field of view of M\,92 and investigate the internal kinematics of the multiple stellar populations. The proper motions are derived by following the recipe by \citet[][see their Section\,5.1]{milone2022b}. 
 We separately reduced each group of images collected at the same epoch through the same filter and camera by using the methods described in Section\,\ref{sec:data} (see Table\,\ref{tab:data} for details on the dataset). To avoid systematic errors in proper motions, we excluded from the analysis the images collected through the F225W, F275W, and F336W filters of UVIS/WFC3 \citep{bellini2011a}.
  We derived the  astrometric and photometric catalog of each group of images and selected the reference frame of the F814W images from GO\,10775 as a master frame. 
   Six-parameter linear equations are used to transform the coordinates of the stars in each catalog into the master frame \citep{anderson2006a}. 
   To derive the proper motion of each star, we plotted the stellar displacements, expressed in milliarcsec, relative to the master frame, against the time in years. We fitted these points with a weighted least-squares straight line and considered the slope as the best estimate of the proper motions \citep[see][for details]{piotto2012a, milone2022b}.
   
   To derive the transformations we used only the bright and unsaturated stars that are well-fitted by the PSF model, according to the criteria of Section\,\ref{sec:data}. Specifically, we calculated proper motions relative to a sample of cluster stars that are selected by using a two-step procedure. 
   We initially selected stars that based on their position on the CMDs are cluster members and derived raw proper motions. Then, we improved the determination of the proper motions by deriving the transformations from those stars that, according to their kinematics, are not cluster members.
   As a consequence of this procedure, the cluster stars have null relative proper motions.

To transform the proper motions from the relative to the absolute scale, we identified the stars for which relative proper motions from {\it HST} and {\it JWST} and absolute proper motions from Gaia DR3 are available. We only considered stars with accurate Gaia DR3 proper motions according to the criteria by \citet{cordoni2020a}, which are based on the proper motion uncertainties and on the
values of the Renormalized Unit Weight Error (RUWE), the astrometric\_gof\_al (As\_gof\_al) parameter.

\section{Multiple stellar populations in M92}\label{sec:MPs}

M\,92 is a well-studied cluster in the context of multiple populations.
The ChM of RGB stars reveals an extended 1G sequence, which hosts 30.4$\pm$1.5\% of cluster stars, and two distinct groups of 2G stars \citep[][]{milone2017a, milone2018a}.  
Recent works based on multi-band {\it HST} photometry indicate that the extended 1G is due to internal metallicity variations of [Fe/H]$\sim$0.15 dex \citep{legnardi2022a} and 2G stars with extreme chemical compositions are enhanced in helium mass fraction by 0.039$\pm$0.006 with respect to the primordial helium abundance (Y=0.246). Moreover, they have enhanced [N/Fe] by $\sim$0.9 dex and are depleted by $\sim$0.5 dex in both [C/Fe] and [O/Fe] \citep[]{meszaros2015a}. Photometric evidence of multiple populations along the AGB is provided by \citet[][]{lagioia2021a}, whereas the detection of multiple populations along the MS is provided by \citet{piotto2015a}, and \citet{nardiello2022a}, who identified a double MS in CMD constructed with the photometry in the F275W filter of UVIS/WFC3 and the F150W NIRCam band.

Further evidence of stellar populations with different chemical compositions among giant stars comes from high-resolution spectroscopy. Star-to-star variations in carbon, oxygen, nitrogen, and sodium are well known since the late seventies \citep[e.g.][and references there in]{sneden1991a, sneden2000a, kraft1994a}. More recently, \citet{meszaros2015a} and \citet{masseron2019a} analyzed the elemental abundances of a large sample of giant stars of M\,92 from the APOGEE survey. 
They detected three main stellar populations with different light-element abundances, including the 1G and two groups of 2G stars. 2G stars with extreme chemical compositions are enhanced in [Al/Fe] and [Si/Fe] by $\sim$1.2 and $\sim$ 0.2 dex, with respect to 1G stars. These stars are also depleted in both [O/Fe] and [Mg/Fe] by $\sim$0.5 dex when compared with the 1G. The 2G stars have intermediate chemical composition, with an aluminum content that is $\sim$0.8 dex higher than that of 1G stars. They are slightly depleted in [O/Fe] and [Mg/Fe] and enhanced in [Si/Fe] by about 0.1 dex.

In the following, we present the photometric diagrams where the multiple populations are more evident, based on the diagrams constructed with the available photometric bands of WFC/ACS, UVIS/WFC3, and NIRCam. Sections\,\ref{subsec:MPsMS} and \ref{subsec:pms} are focused on bright MS stars and their kinematics, whereas Section\,\ref{subsec:MPsMD} is dedicated to the M-dwarfs.

\subsection{Multiple populations along the main sequence}\label{subsec:MPsMS}
\begin{figure*}
    \centering
    \includegraphics[width=15.5cm]{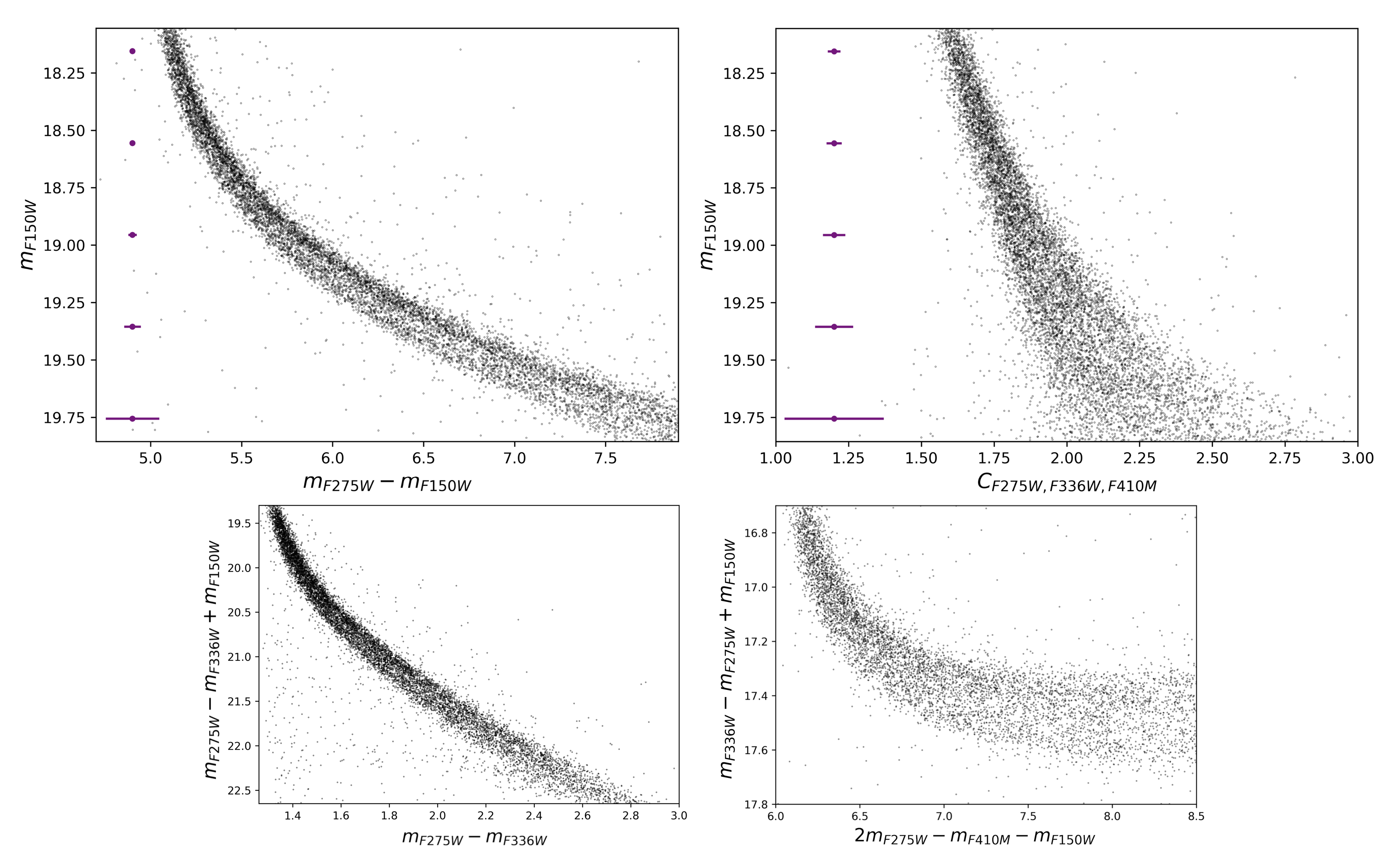}
    \caption{Top panels: $m_{\rm F150W}$ vs.\,$m_{\rm F275W}-m_{\rm F150W}$ (left) and $m_{\rm F150W}$ vs.\,$C_{\rm F275W,F336W,F410M}$ diagram (right) of MS stars. Bottom panels: $m_{\rm F275W} - m_{\rm F336W} + m_{\rm F150W}$ vs.\,$m_{\rm F275W}-m_{\rm F336W}$ (left) and $m_{\rm F336W} - m_{\rm F275W} + m_{\rm F150W}$ vs.\,$2m_{\rm F275W}-m_{\rm F410M} - m_{\rm F150W}$ (right) diagrams.}
    \label{fig:cmds}
\end{figure*}

 The CMD constructed with photometry in the F275W band of UVIS/WFC3 and the F150W NIRCam band clearly shows a split MS  and similar results are obtained by using the F814W filter of WFC3/ACS or the available NIRCam filters \citep[see also][]{nardiello2022a}. 
 As shown in the upper-left panel of Figure \ref{fig:cmds}, the two MSs are nearly mixed around the turn-off and the $m_{\rm F275W}-m_{\rm F150W}$ color separation between the blue MS and the red MS increases towards fainter magnitudes. 

 The $C_{\rm F275W,F336W,F410M}$ pseudo-color is another efficient tool to identify multiple stellar populations in GCs \citep{milone2013a}. As shown in the upper-right panel of Figure \ref{fig:cmds}, where we plot $m_{\rm F150W}$ against $C_{\rm F275W,F336W,F410M}$ pseudo-CMD, the MS is intrinsically broadened.

 Evidence of multiple populations along the MS is provided by the pseudo-CMDs shown in the bottom panels of Figure \ref{fig:cmds}, namely $m_{\rm F275W}-m_{\rm F336W}+m_{\rm F150W}$ vs.\,$m_{\rm F275W}-m_{\rm F336W}$ and $m_{\rm F336W}-m_{\rm F275W}+m_{\rm F150W}$ vs.\,$2 m_{\rm F275W}-m_{\rm F410M}-m_{\rm F150W}$, which are similar to the diagrams introduced by \citet{milone2015a} for NGC\,2808 but use the F410M and F150W bands instead of F438W and F814W.

\begin{figure*}
    \centering
    \includegraphics[width=14.5cm]{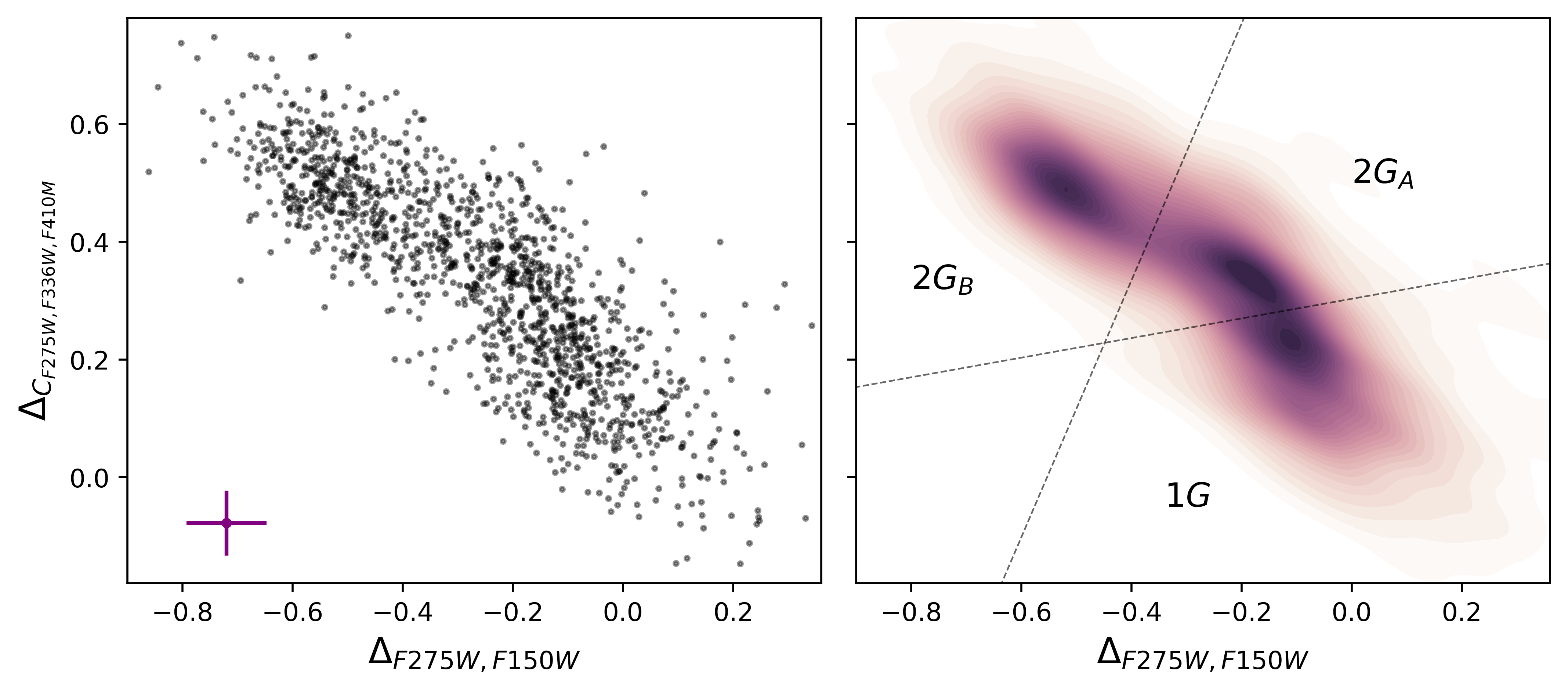}
    \caption{$\Delta_{C \rm F275W,F336W,F410M}$ vs.\,$\Delta_{\rm F275W,F150W}$ ChM for MS stars with $19.26<m_{\rm F150W}<19.56$ mag (left panel). The corresponding Hess diagram is plotted in the right panel. The dotted lines separate the bulk of 1G, 2G$_{\rm A}$, and 2G$_{\rm B}$ stars.}
    \label{fig:chm1}
\end{figure*}

To better identify multiple populations along the MS, we used the photometric diagrams plotted in the upper panels of Figure \ref{fig:cmds} to construct the $\Delta_{C \rm F275W,F336W,F410M}$ vs.\,$\Delta_{\rm F275W,F150W}$ ChM of MS stars, following the procedure introduced by \citet[][]{milone2017a}. 
 Results are shown in Figure \ref{fig:chm1}, where we show the ChM and the corresponding Hess diagram for stars in the MS region with $19.26<m_{\rm F150W}<19.56$ mag, where multiple populations are more clearly visible.

 The bulk of stars near the origin of the ChM corresponds to the 1G, whereas 2G stars are extended towards larger values of $\Delta_{C \rm F275W,F336W,F410M}$ and smaller values of $\Delta_{\rm F275W,F150W}$. Clearly, 2G stars are distributed along an extended sequence, and define two main stellar overdensities around $\Delta_{C \rm F275W,F336W,F410M} \sim 0.25$ and 0.50 mag, which we called 2G$_{\rm A}$ and 2G$_{\rm B}$, respectively. The dashed lines that we determined empirically in the right panel of Figure \ref{fig:chm1} identify three stellar groups that are mostly populated by 1G, 2G$_{\rm A}$, and 2G$_{\rm B}$ stars.

 The small separation between 1G and 2G$_{\rm A}$ stars in the ChM indicates that they share similar chemical compositions. In the context of the formation scenario proposed by \citet{renzini2022a}, this evidence suggests that 1G stars did not have sufficient time to significantly pollute the interstellar medium of the cluster before the 2G$_{\rm A}$ population formed, indicating that 2G$_{\rm A}$ stars formed shortly after 1G. Conversely, the 2G$_{\rm B}$ population is clearly separated from the other populations, indicating a separate star formation event.

\subsection{Internal kinematics of multiple stellar populations}\label{subsec:pms}
 To investigate the kinematics of the distinct stellar populations of M\,92, we analyzed the proper motion dispersion of 1G, 2G$_{\rm A}$, and 2G$_{\rm B}$ stars identified in the previous section at different radial distances from the cluster center.
 We divided the field of view into various circular annuli by using the method of the naive estimator \citep{silverman1986a}. Specifically, we defined a series of six points in the interval between 19 and 92 arcsec from the center of M\,92 separated by a distance
of $d$=24 arcsec. The bins are defined over a grid of points, which are separated by steps of $d$/2 in distance. 

For each radial bin, we calculated the proper motion velocity dispersion along the radial and the tangential directions ($\sigma_{\rm R}$ and $\sigma_{\rm T}$). To do that, we extended to M\,92 the procedure based on maximum-likelihood by \citet{mackey2013a} and \citet{marino2014a}. 
 We assumed that the stellar proper motions have a normal distribution, which is described by their average value and intrinsic dispersion. The observed proper motion distribution is also affected by measurement uncertainties. 
 The intrinsic dispersion is inferred through the maximization of the logarithm of the joint probability function for the observed proper motions. 
  We estimated the errors associated with the dispersion determinations  by means of bootstrapping with replacements performed 1,000 times. We considered the 68.27$^{\rm th}$ percentile of the bootstrapped measurements as the best estimate of the proper motion dispersion uncertainty.
  For completeness, we extended the procedure above to all MS stars with $m_{\rm F814W}<20$ mag. 

\begin{centering} 
\begin{figure*} 
\centering
  \includegraphics[width=14.5cm,clip]{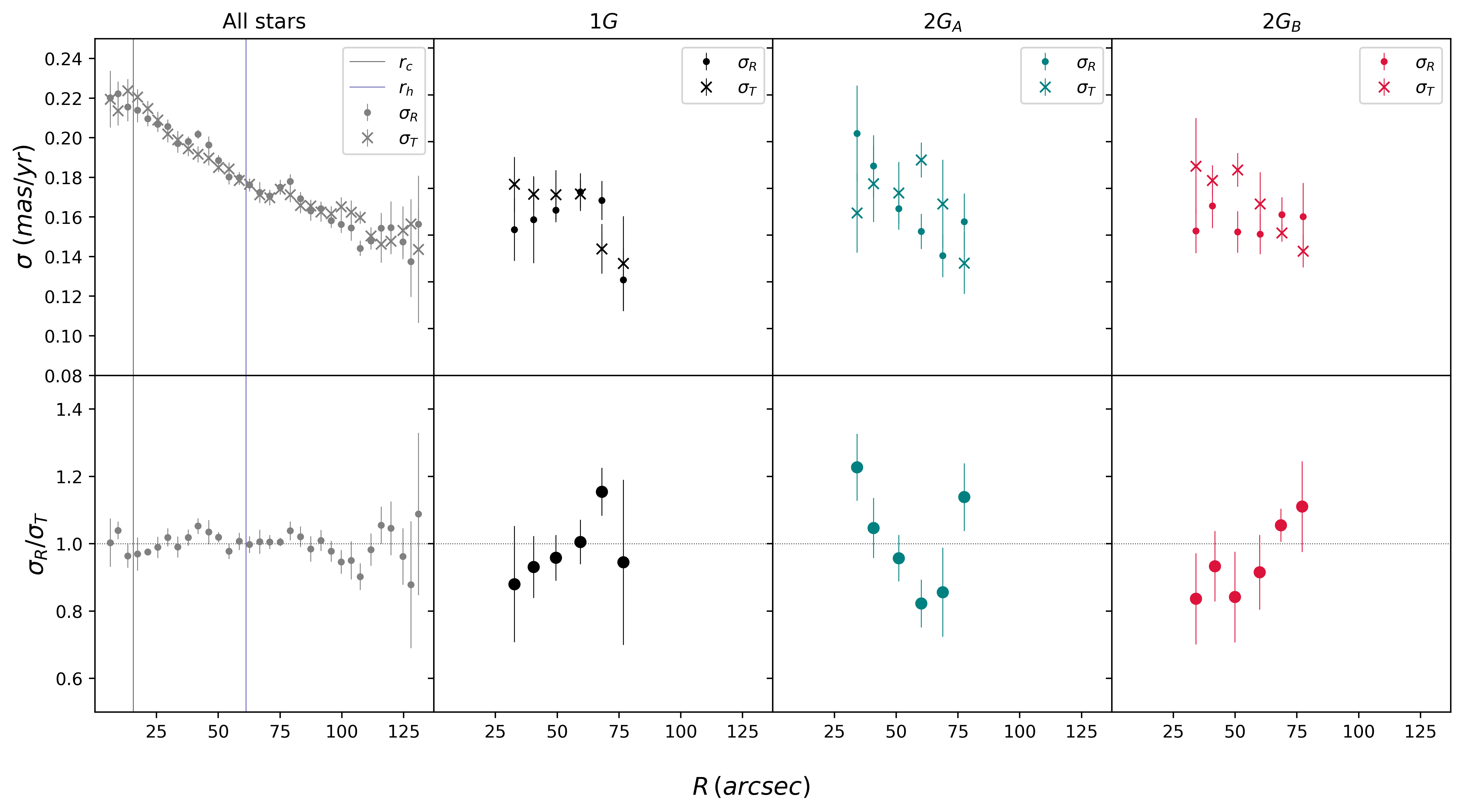}
 \caption{Velocity dispersion along the tangential (cross symbols) and radial directions (filled circles) as a function of the radial distance from the cluster center (top panels). Tangential to radial isotropy against the radial distance from the cluster center (bottom panels). We show results for all  MS stars with $m_{\rm F814W}<$ 20.0, for 1G stars, 2G$_{\rm A}$, and 2G$_{\rm B}$ stars. The two vertical lines in the left panels indicate the core radius ($r_c$) and the half-light radius ($r_h$) of the cluster, respectively.}
 \label{fig:pm2} 
\end{figure*} 
\end{centering}

As shown in the upper-left panel of Figure \ref{fig:pm2}, the proper motion dispersion of all M\,92 stars ranges from $\sim$0.20 mas/yr near the cluster center to $\sim$0.15 mas/yr at a radial distance of $\sim$140 arcsec, which corresponds to $\sim$2.3 half-light radii.
1G, 2G$_{\rm A}$, and 2G$_{\rm B}$ stars (black, green, and red points in the upper panels) share the same  velocity dispersion distributions along both the tangential and radial direction. The radial interval covered by the identified 1G, 2G$_{\rm A}$, and 2G$_{\rm B}$ stars is $\sim$0.6-1.3 half-light radii. This radial interval is smaller than the one analyzed for all stars because it corresponds to the field where JWST and HST observations overlap.
We find isotropic motions in each stellar population as demonstrated by the fact that the ratio between $\sigma_{\rm T}$ and $\sigma_{\rm R}$ (lower panels of Figure \ref{fig:pm2}) is consistent with being $\sim$1 in the analyzed radial interval. 

In the past decade, work based on high-precision proper motion, mostly from {\it HST} multi-epoch images and from Gaia data, allowed the astronomy to investigate the internal kinematics of stellar populations in GCs \citep[e.g.\,][]{richer2013a, libralato2023a}.
 While the 2G stars of some massive and dinamically young GCs, like $\omega$\,Centauri, NGC\,2808 and 47\,Tucanae exhibit more radially anisotropic velocity distributions than the 1G \citep[e.g.\,][]{bellini2015a, bellini2017a, milone2018b, cordoni2020b}, both 1G and 2G stars of other clusters exhibit nearly isotropic velocity distributions \citep[e.g.][]{libralato2019a, cordoni2020a}.
 In a recent work, \citet{libralato2022a, libralato2023a} derived the internal proper motions of 56 GCs and their stellar populations, by using multi-epoch {\it HST} images. 
 Our results on M\,92 stars are consistent with the results by Libralato and collaborators, who concluded that, similarly to the most dynamically evolved GCs, M\,92 exhibits nearly isotropic motions for radii smaller than $\sim$2.5 times the half-light radius. 
 
\subsection{Multiple populations along M-dwarfs}\label{subsec:MPsMD}
\begin{figure*}
    \centering
    \includegraphics[height=9.85cm,trim={0.3cm 0.5cm 1.1cm 0.0cm},clip]{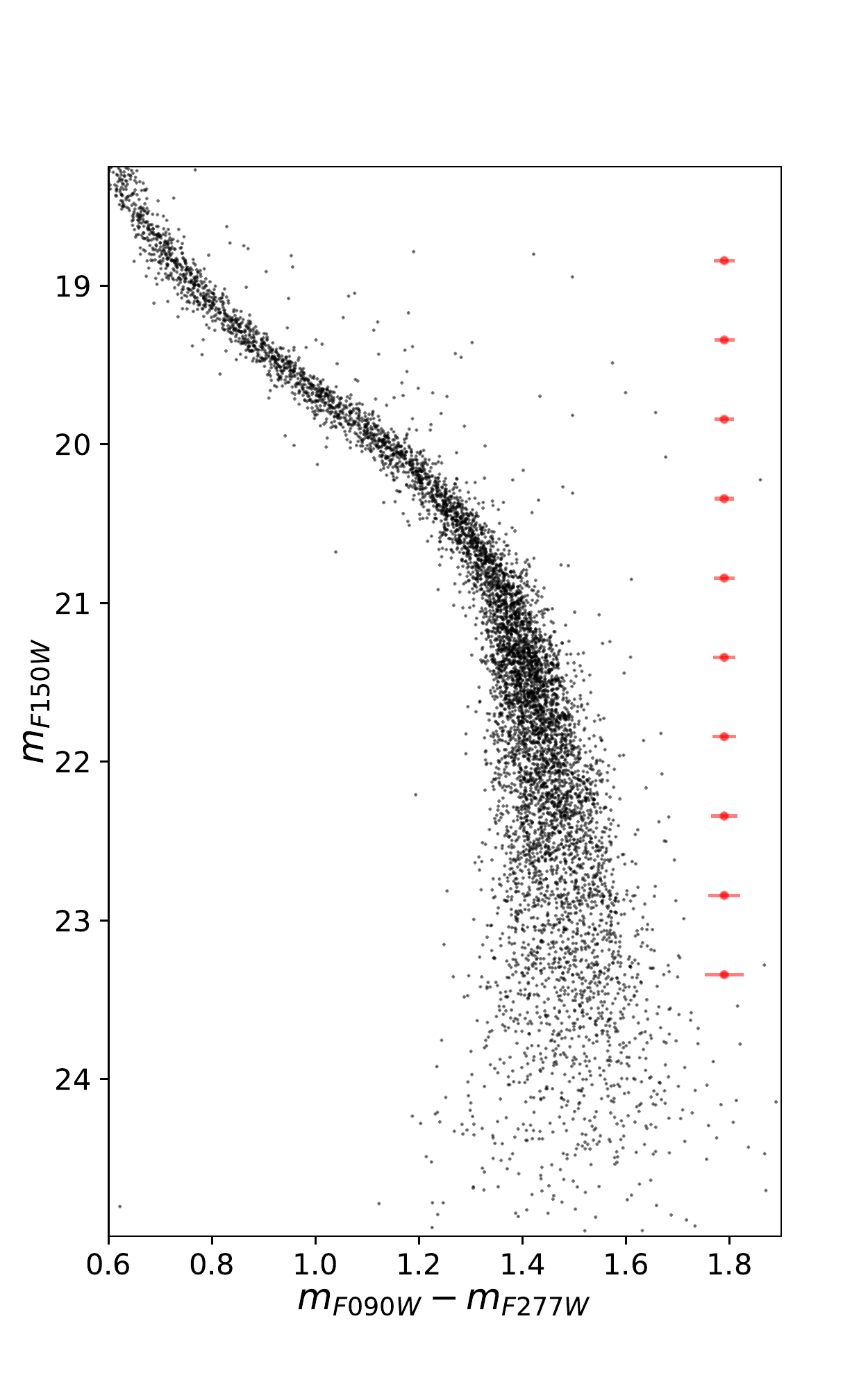}
    \includegraphics[height=9.85cm, trim={1.5cm 0.5cm 0.5cm 0.0cm},clip]{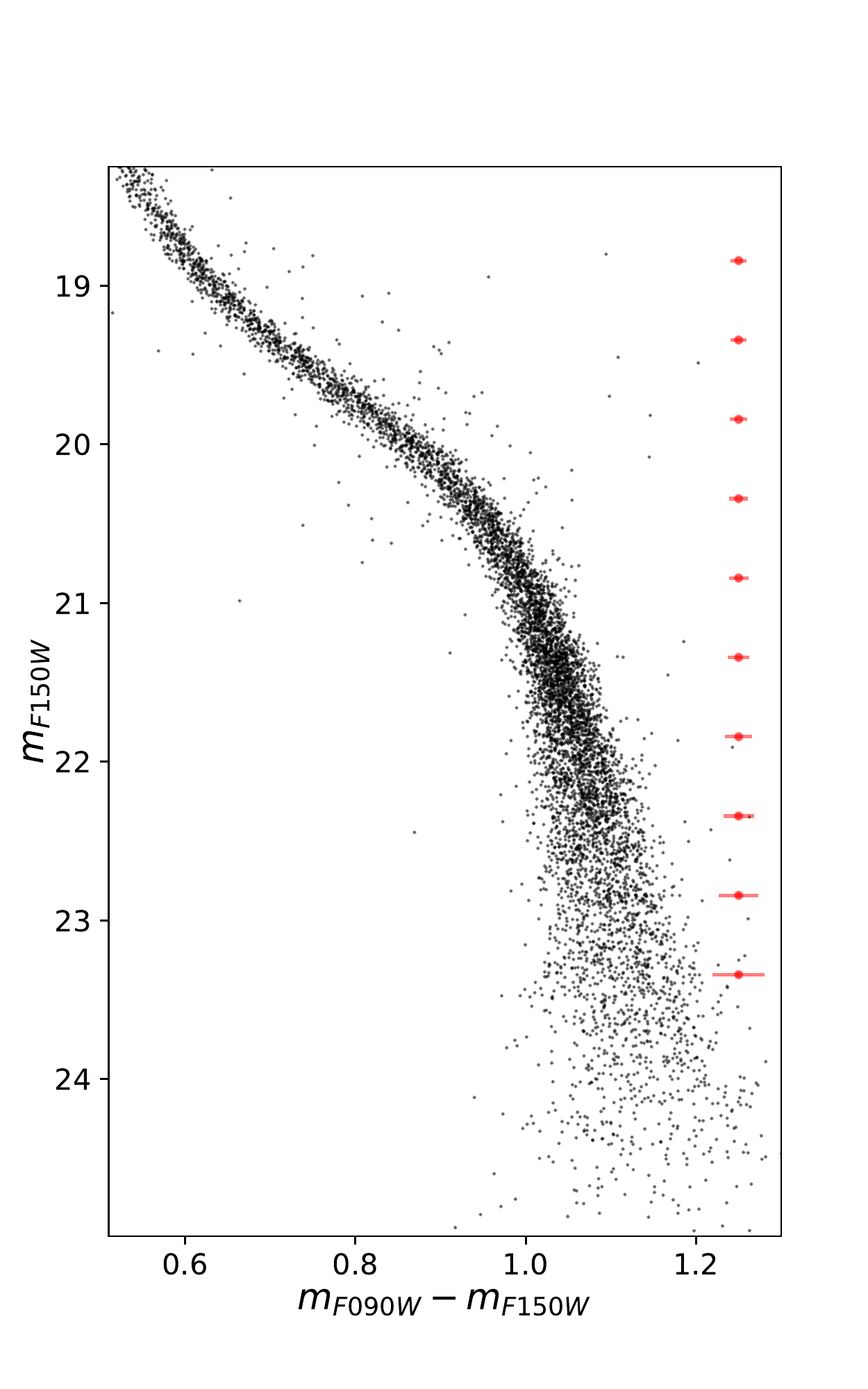}
    \includegraphics[width=5.5cm,clip]{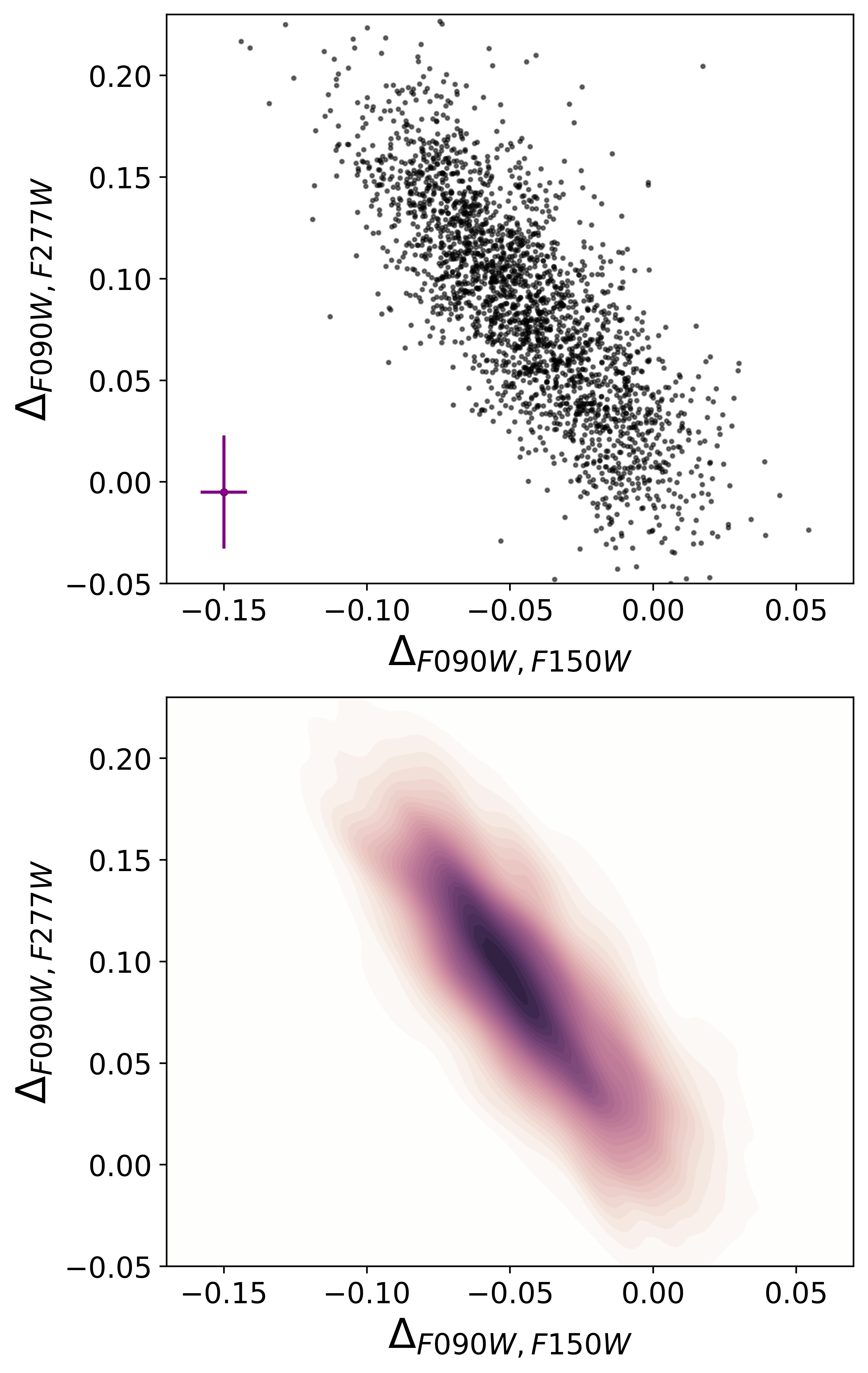}
    \caption{$m_{\rm F150W}$ vs.\,$m_{\rm F090W}-m_{\rm F277W}$ (left) and  $m_{\rm F150W}$ vs.\,$m_{\rm F090W}-m_{\rm F150W}$ (middle) CMDs of stars with radial distance from the M\,92 center larger than 1.5 arcmin. The right panels show the ChM for M-dwarfs (top) and the corresponding Hess diagram (bottom).}
    \label{fig:cmd2}
\end{figure*}

The CMDs constructed with the F090W, F150W, and F277W filters of NIRCam reveal a broad MS below the knee. As illustrated in Figure \ref{fig:cmd2}, the color spread of M-dwarfs is significantly wider than what is expected from the photometric errors alone, thus indicating the presence of multiple stellar populations among low-mass stars. 

 We used the $m_{\rm F150W}$ vs.\,$m_{\rm F090W}-m_{\rm F277W}$ and  $m_{\rm F150W}$ vs.\,$m_{\rm F090W}-m_{\rm F150W}$ CMDs to construct the  $\Delta_{\rm F090W,F277W}$ vs.\,$\Delta_{\rm F090W,F150W}$ ChM plotted in the top-right panel of Figure \ref{fig:cmd2}. We only considered M-dwarfs with 19.26 $<m_{\rm F150W}<$ 19.56 mag (top), which is the MS region where the color broadening is more evident.
  As suggested by the Hess diagram (bottom-right panel of Figure \ref{fig:cmd2}) the M-dwarfs exhibit a continuous distribution in the ChM, without any clear separation between 1G and 2G stars.

\section{Comparison with synthetic model atmospheres}\label{sec:teo}
To investigate the behavior of multiple stellar populations in photometric diagrams constructed with the {\it HST} filters, we derived the colors and magnitudes of isochrones that account for the chemical compositions of 1G and 2G stars.
 Similar to what we did in previous papers \citep[e.g.\,][]{milone2018a, milone2023a}, we selected fifteen points along the isochrone that provides the best fit with the $m_{\rm F606W}$ vs.\,$m_{\rm F606W}-m_{\rm F814W}$ CMD of M\,92 and extracted the effective temperature, $T_{\rm eff}$, and gravity, $g$, of each point.
 We used the isochrones from the Dartmouth database \citep{dotter2008a} with iron abundance [Fe/H]=$-$2.4\,dex and [$\alpha$/Fe]=0.4\,dex and adopted the age of 13.0\,Gyr, distance modulus ($m-M$)$_{0}$=14.75 mag and reddening E(B$-$V)=0.03 mag, which are similar to the values derived by \cite{dotter2010a}. 
 We calculated models with primordial helium content, Y=0.246, and enhanced helium abundances, Y=0.33.

For each pair of stellar parameters, we computed a stellar atmosphere structure with ATLAS\,12, which is the model atmosphere code developed by Robert Kurucz \citep[e.g.\,][]{kurucz1970a, kurucz1993a} and ported to Linux by \citet{sbordone2004a}. 
We computed a reference synthetic spectrum with similar chemical composition as a 1G star (i.e.\, solar-scaled abundances of C and N, and [O/Fe]=0.40), and a comparison spectrum with [C/Fe]=$-$0.5 dex, [O/Fe]=$-$0.1 dex, [N/Fe]=1.2 dex, which is representative of a 2G star. 

\begin{centering} 
\begin{figure*} 
\centering
  \includegraphics[height=11.cm,trim={0.0cm 5cm 6.2cm 3.0cm},clip]{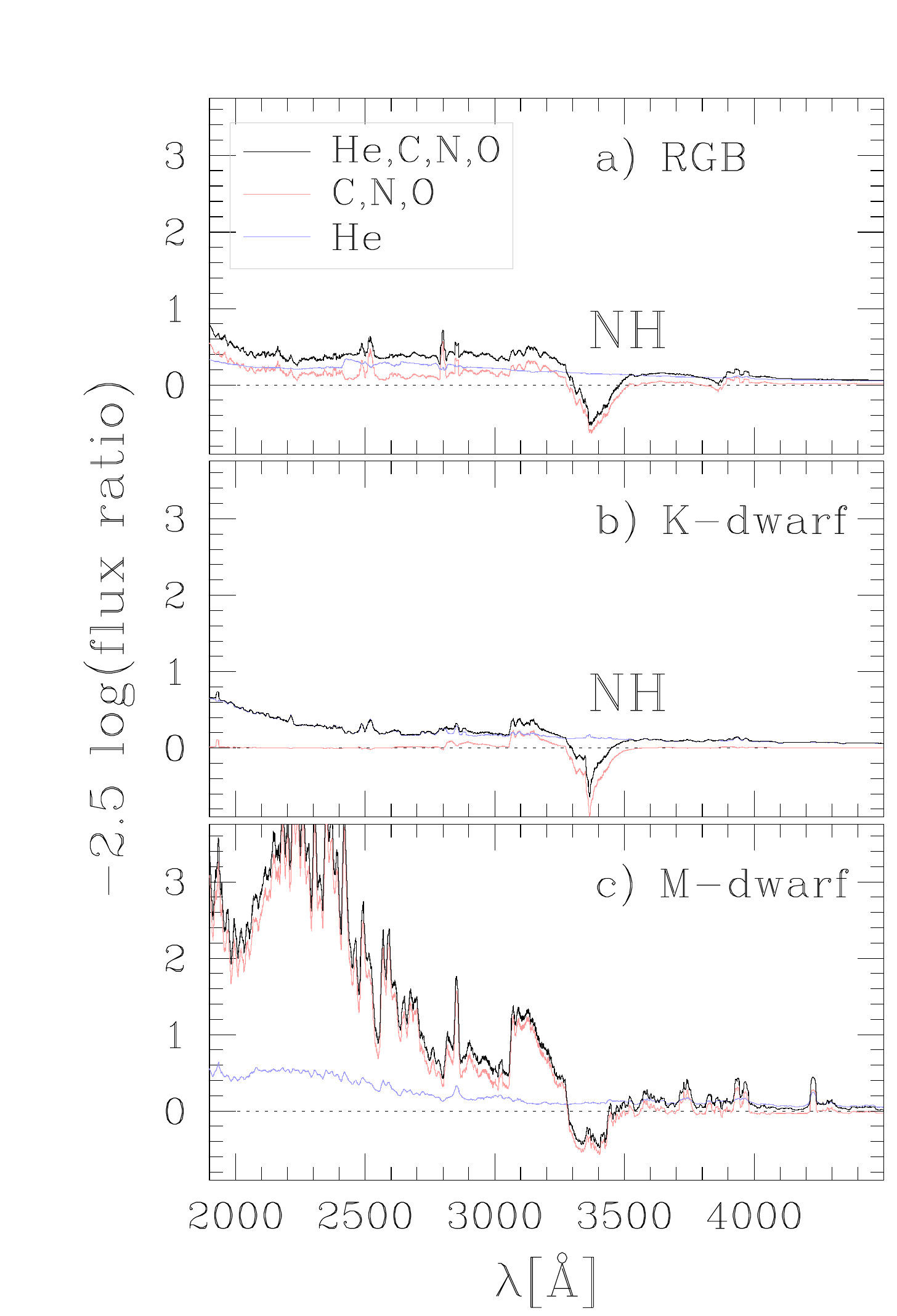}
  \includegraphics[height=11.cm,trim={1.5cm 5cm 0cm 3.0cm},clip]{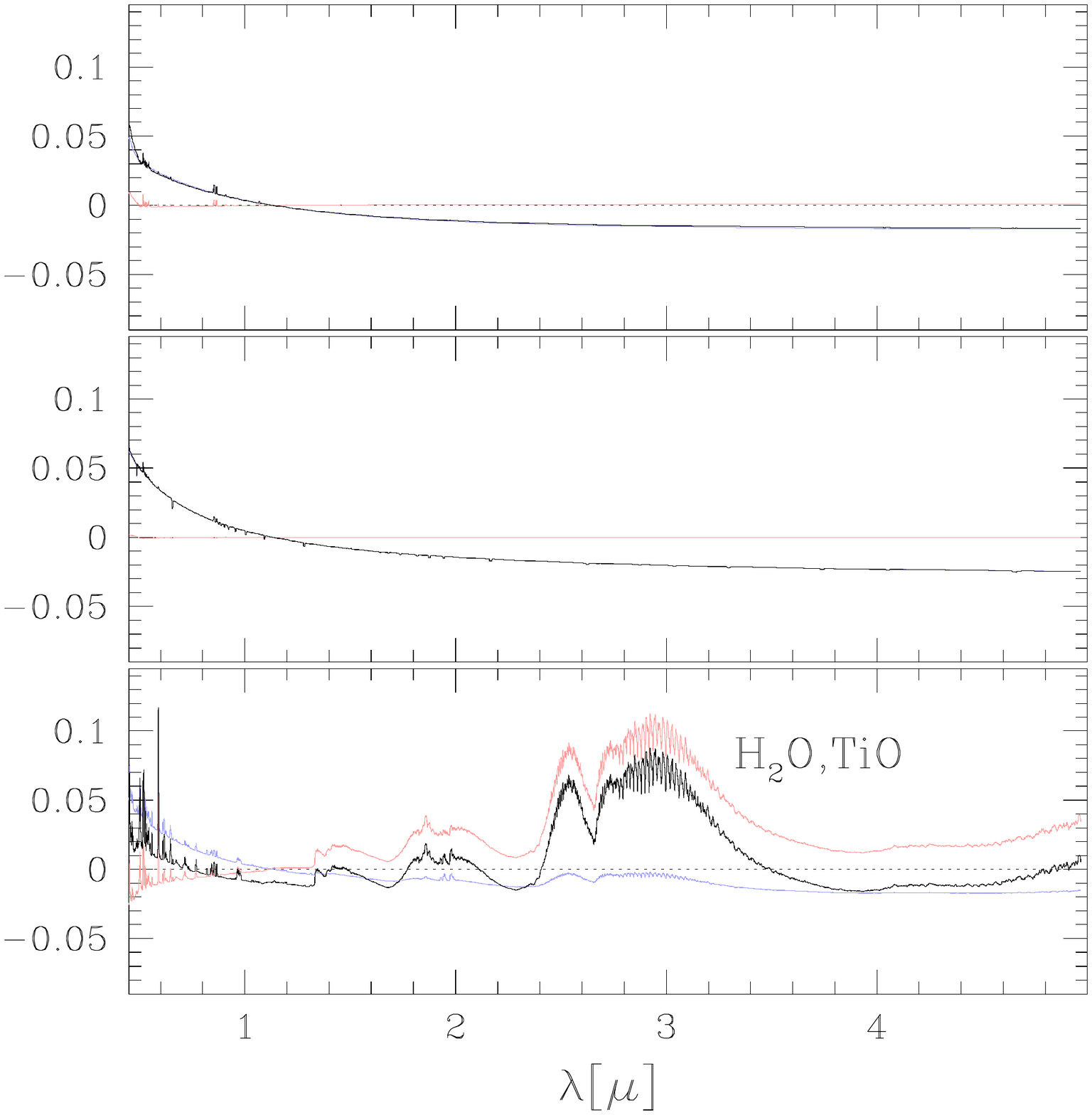}
 \caption{Flux ratio between the simulated spectra of stars with 2G-like chemical composition and the spectra of a 1G star (black lines). The pink lines correspond to spectra with the same helium content as 1G stars but different C, N, and O abundances. The blue lines are obtained from helium-enhanced stars with the same C, N, and O content as the 1G. Panels a, b, and c refer to RGB, upper-MS, and M-dwarf stars, respectively. We indicate the NH molecule which is the main responsible for the F336W magnitude difference of 2G and 1G RGB and MS stars, and the H2O and TiO molecules that contribute to the F277W flux difference of multiple populations among M-dwarfs.}
 \label{fig:FigDF} 
\end{figure*} 
\end{centering}

The synthetic spectra are constructed with the computer program SYNTHE \citep{kurucz1981a, castelli2005a, sbordone2007a} over the wavelength interval covered by UVIS/WFC3, WFC/ACS, and NIRCam filters, between 2,000 and 52,000 \AA. 
As an example, the black lines of Figure\,\ref{fig:FigDF} represent the logarithm of the flux ratio between  He-rich stars with 2G-like C, N, and O abundances and 1G stars at the same F115W magnitude. 
 To display the effect of changing helium or C, N, O alone, we show the flux ratio obtained from a He-rich star with the same C, N, and O abundances as the 1G (blue line) and the flux ratio derived from a star with 2G-like abundances of C, N, and O but primordial helium content (Y=0.246, pink line). The panels a correspond to RGB stars with $M_{\rm F115W}=-3.80$ mag, whereas panels b and c refer to MS stars with $M_{\rm F115W}=5.14$ and 8.29 mag, respectively.
 Clearly, in the wavelength interval covered by NIRCam, flux of the spectra of bright MS and RGB stars, is nearly insensitive to the adopted C, N, and O variations, as demonstrated by the fact that pink lines are close to zero. On the contrary, the fluxes of M-dwarfs spectra are significantly affected by the light-element content. The most pronounced difference between 1G and 2G spectra occurs around 2.5-3.2 $\mu$m, where 2G stars exhibit higher fluxes than 1G stars with the same F115W magnitude. Other significant flux differences occur around $\lambda \sim 1.6$ $\mu$m, $\lambda \sim 2.0$ $\mu$m, and $\lambda \gtrsim 4.6$ $\mu$m.

\begin{centering} 
\begin{figure*} 
\centering
    \includegraphics[height=5.5cm,trim={0.6cm 6.8cm 0.0cm 10.5cm},clip]{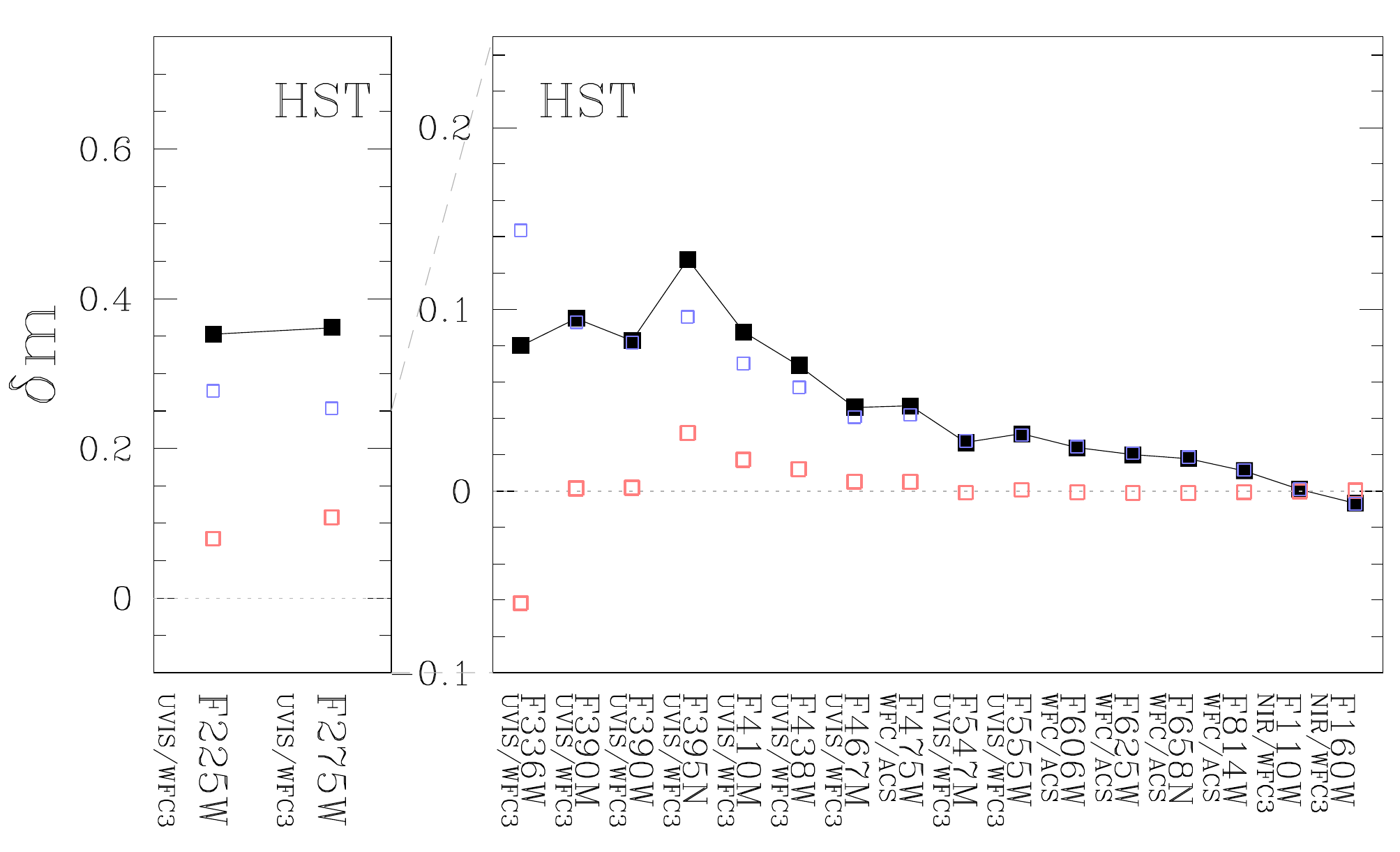}
  \includegraphics[height=5.5cm,trim={1.85cm 6.8cm 0.0cm 10.5cm},clip]{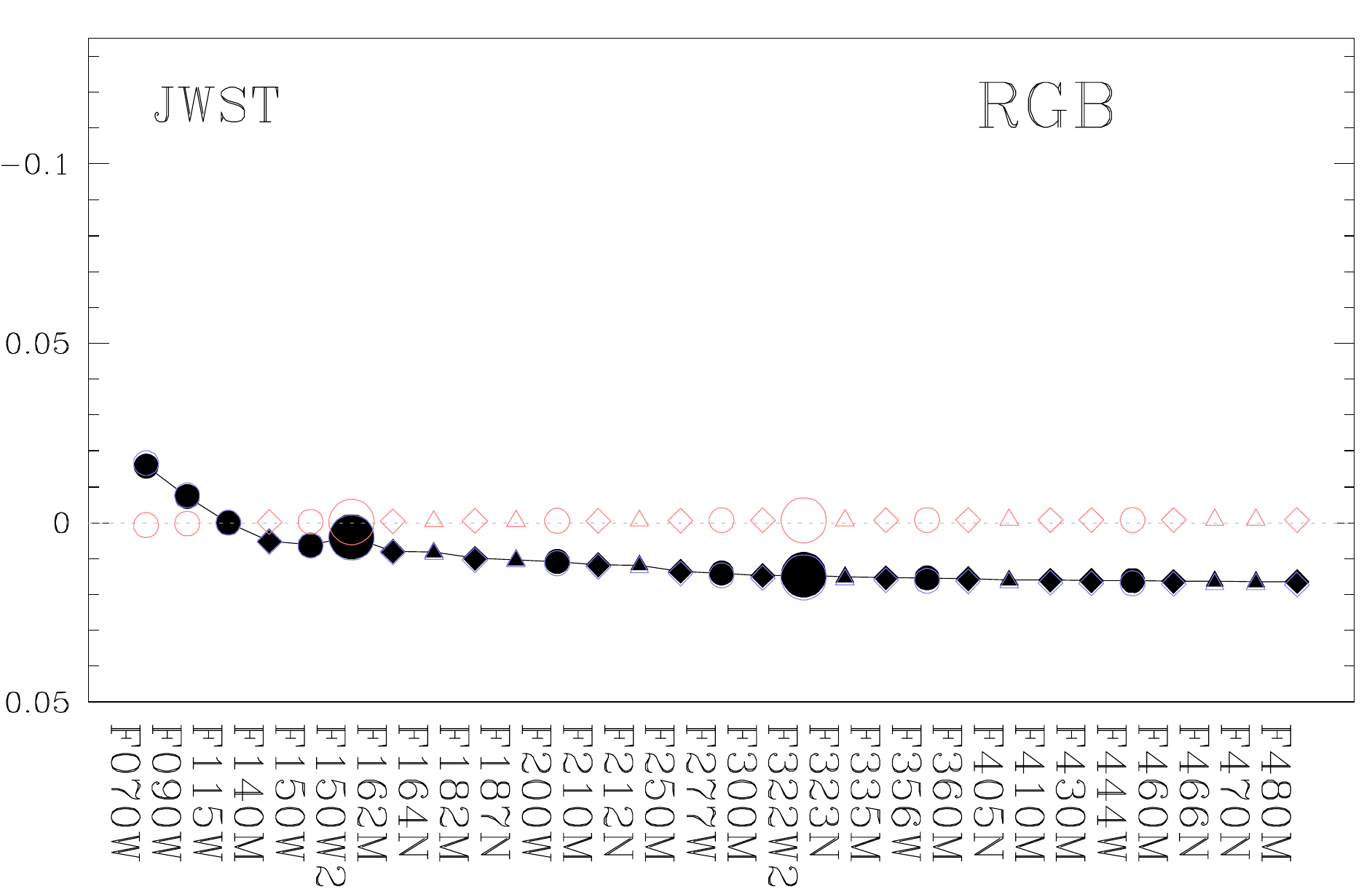}
      \includegraphics[height=5.5cm,trim={0.6cm 6.8cm 0.0cm 10.5cm},clip]{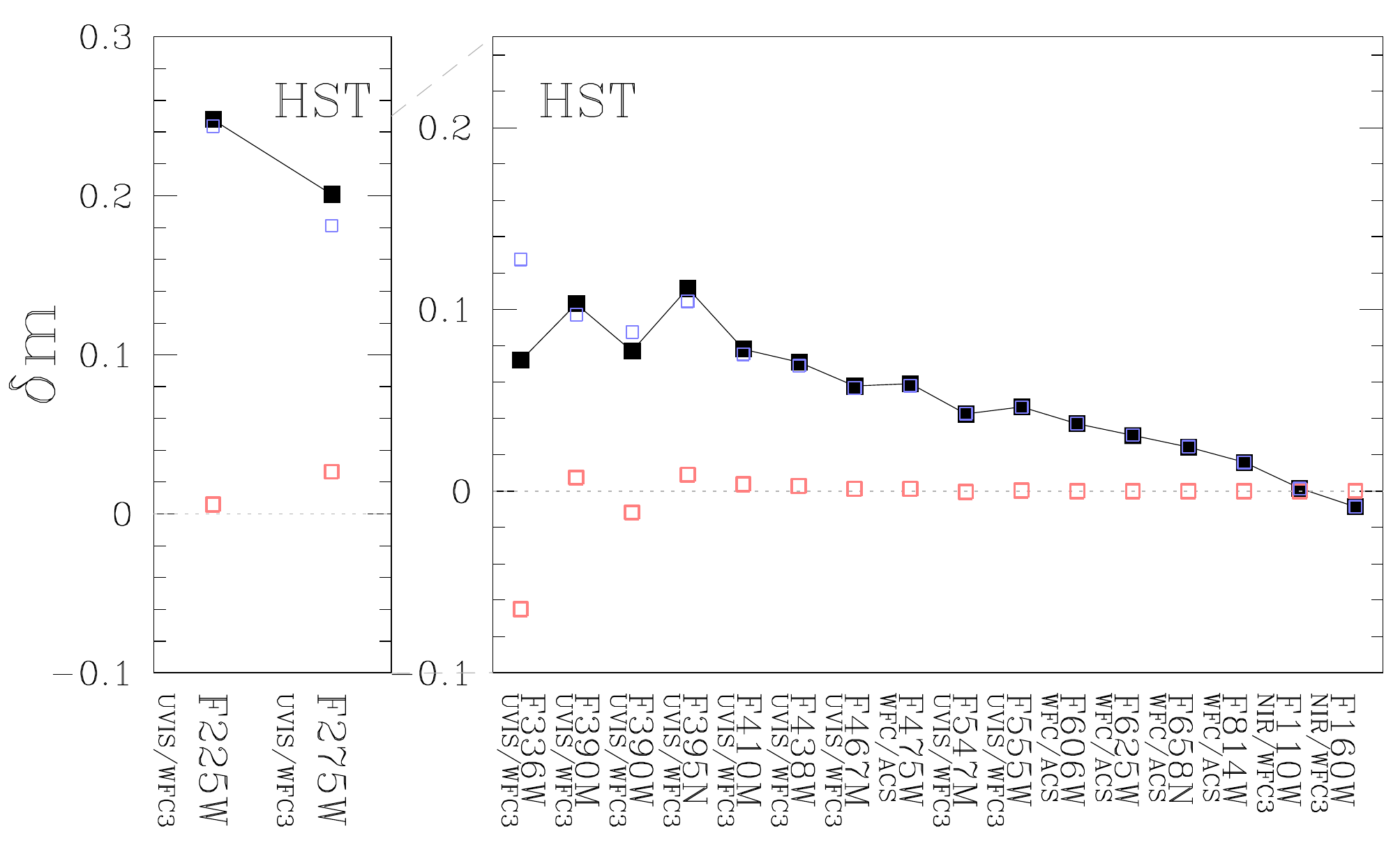}
  \includegraphics[height=5.5cm,trim={1.85cm 6.8cm 0.0cm 10.5cm},clip]{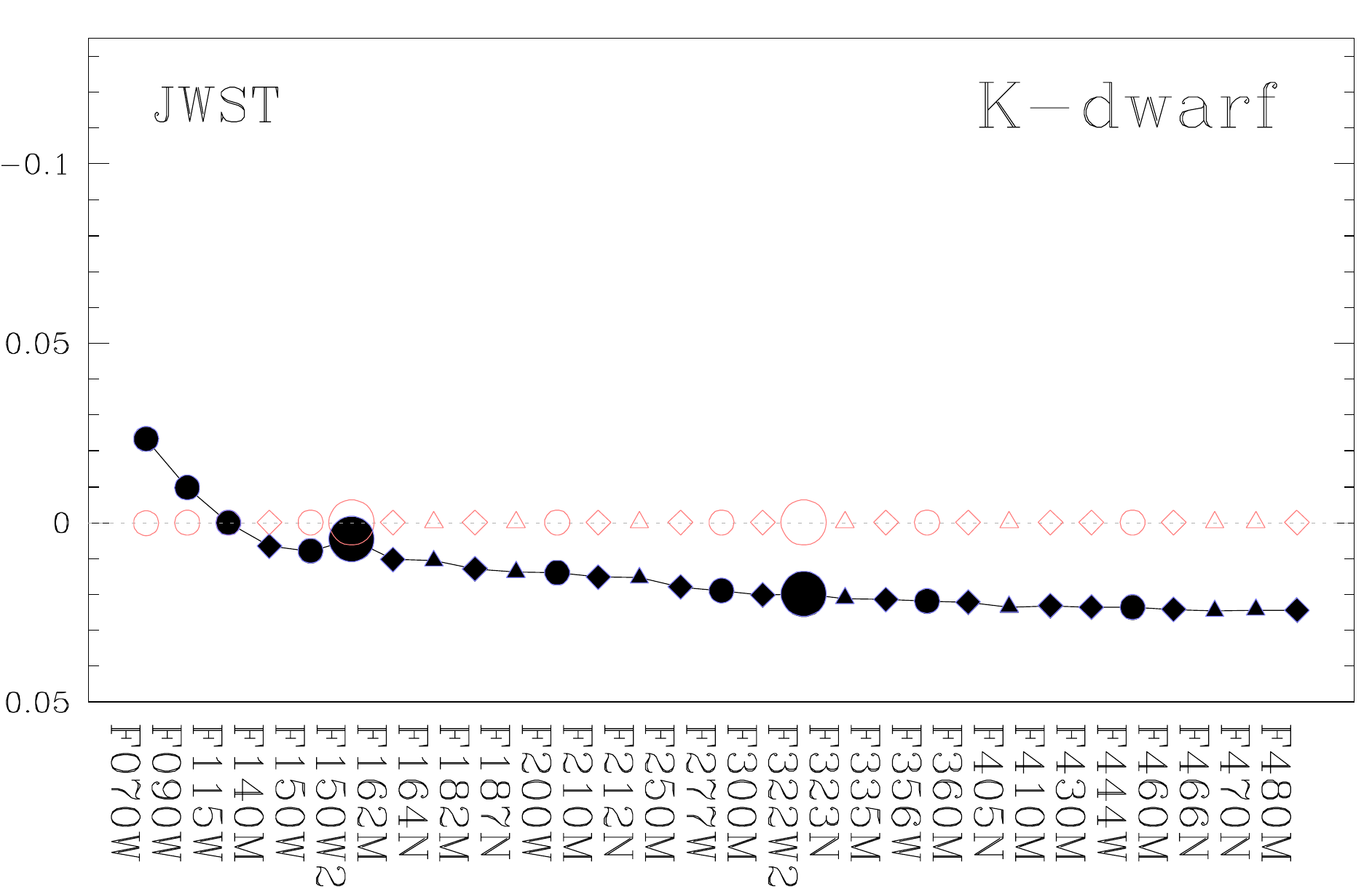}
       \includegraphics[height=5.5cm,trim={0.6cm 6.8cm 0.0cm 10.5cm},clip]{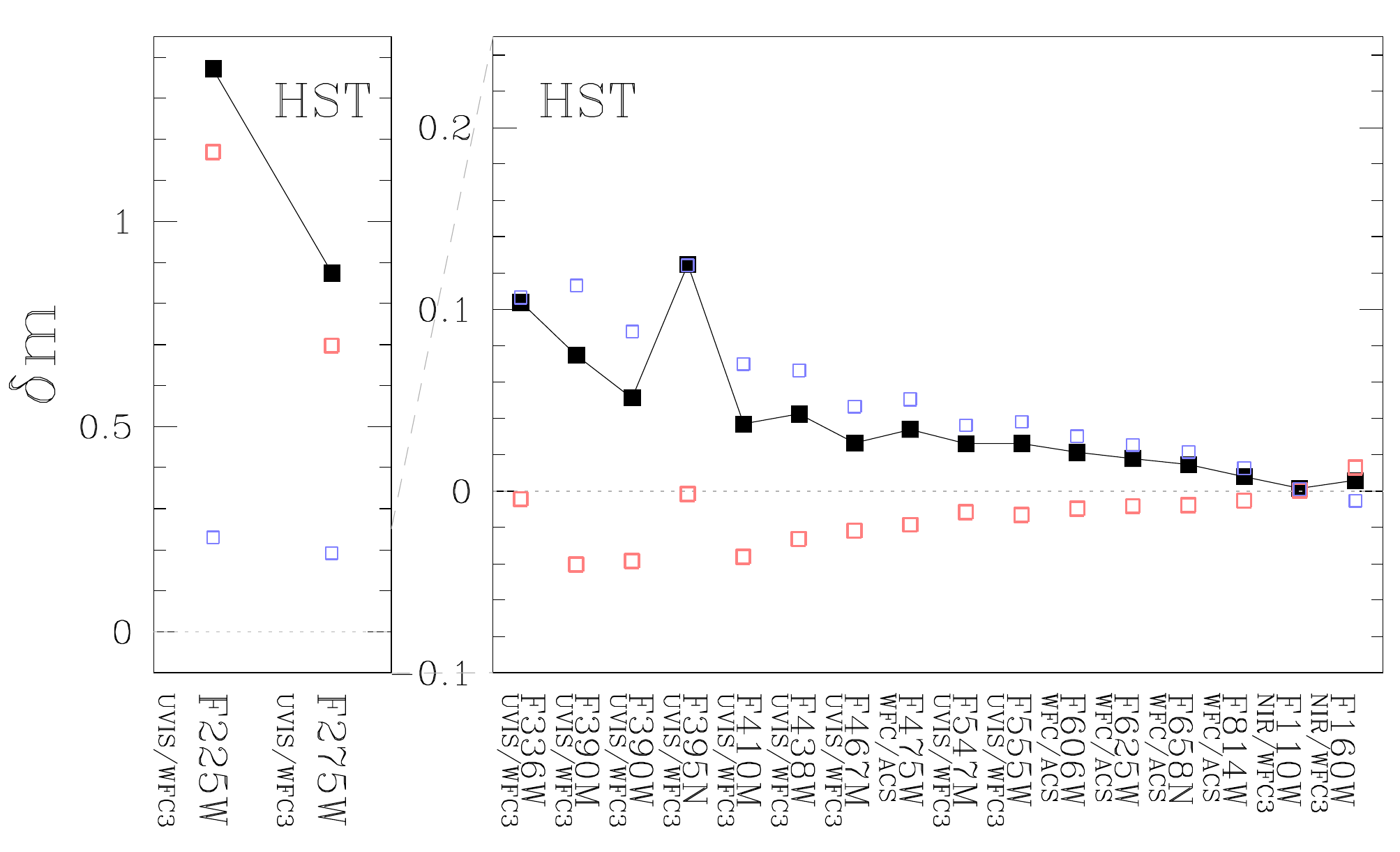} 
  \includegraphics[height=5.5cm,trim={1.85cm 6.8cm 0.0cm 10.5cm},clip]{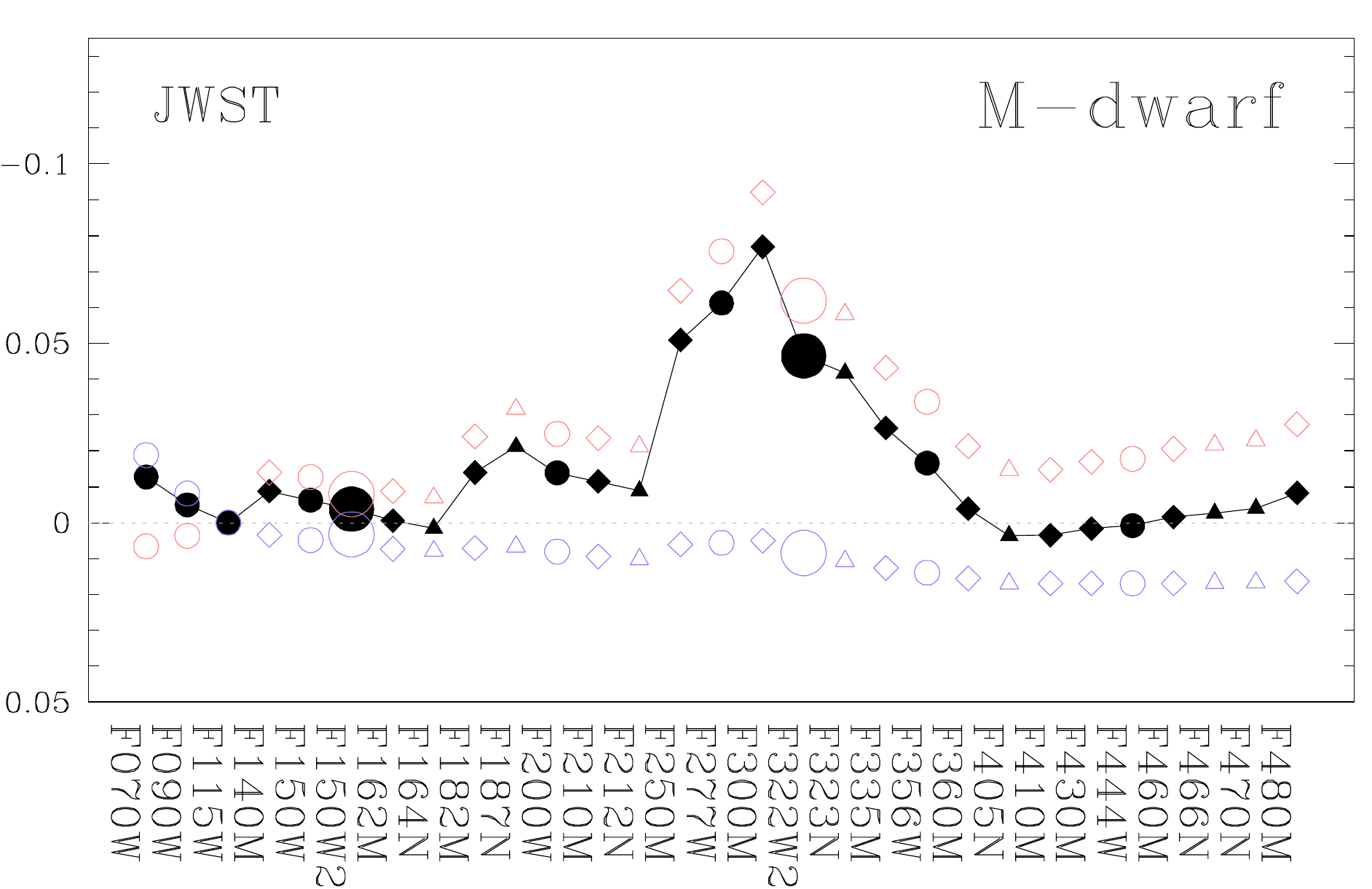}
 \caption{
 Magnitude difference derived from simulated spectra of Figure \ref{fig:FigDF}. Black symbols correspond to stars with 1G-like and 2G-like abundances of He, C, N, and O. 
 The pink points refer to stars with the same helium content but different abundances of C, N, and O, while the blue ones are derived from spectra with the same C, N, and O content as 1G stars but enhanced helium. 
 Results for RGB stars, K-dwarfs, and M-dwarfs are plotted in the top, middle, and bottom panels, respectively. 
 The narrow, middle, wide, and extra-wide passbands  NIRCam filters are represented with triangles, diamonds, circles, and large circles.  }
 \label{fig:DMAG} 
\end{figure*} 
\end{centering} 

The simulated spectra are integrated over the transmission curves of the NIRCam,  WFC/ACS, and UVIS/WFC3 filters that are available for M\,92. 
We calculated the magnitude differences, $\delta m_{\rm X}$, between each comparison spectrum and the  reference spectrum. 
Results are illustrated in Figure \ref{fig:DMAG} for the RGB, M-dwarf, and K-dwarf stars analyzed in Figure \ref{fig:FigDF}. 

 Helium variations are mostly responsible for magnitude differences between 2G and 1G stars in K-dwarfs and RGB stars in NIRCam and {\it HST} filters. 
 The $\delta_{\rm m}$ value associated with helium variations monotonically decreases with the filter central wavelength, but the magnitude separation between 2G and 1G stars is much smaller in NIR filters than in optical and UV bands. 
  The abundances of nitrogen and oxygen mostly affect the UV spectral region. In particular, the OH molecular bands affect the F225W and F275W flux, while NH molecules are responsible for F336W magnitude differences. 
  
 The most pronounced flux difference between 2G and 1G M-dwarfs in the NIR involves the F250M, F277W, F300M, F322W2, and F323N bands and are associated with molecules composed of oxygen atoms (mostly H$_2$O and TiO). We find negligible magnitude variations in the F070W filter of NIRCam and the F606W filter of WFC/ACS, in contrast with what is observed among M-dwarfs with [Fe/H]=$-$0.75, where there are large magnitude differences between 1G and 2G stars in these filters \citep{milone2023a}.

\begin{centering} 
\begin{figure*} 
\centering
    \includegraphics[height=7.6cm,trim={0.6cm 5cm 4.7cm 4.5cm},clip]{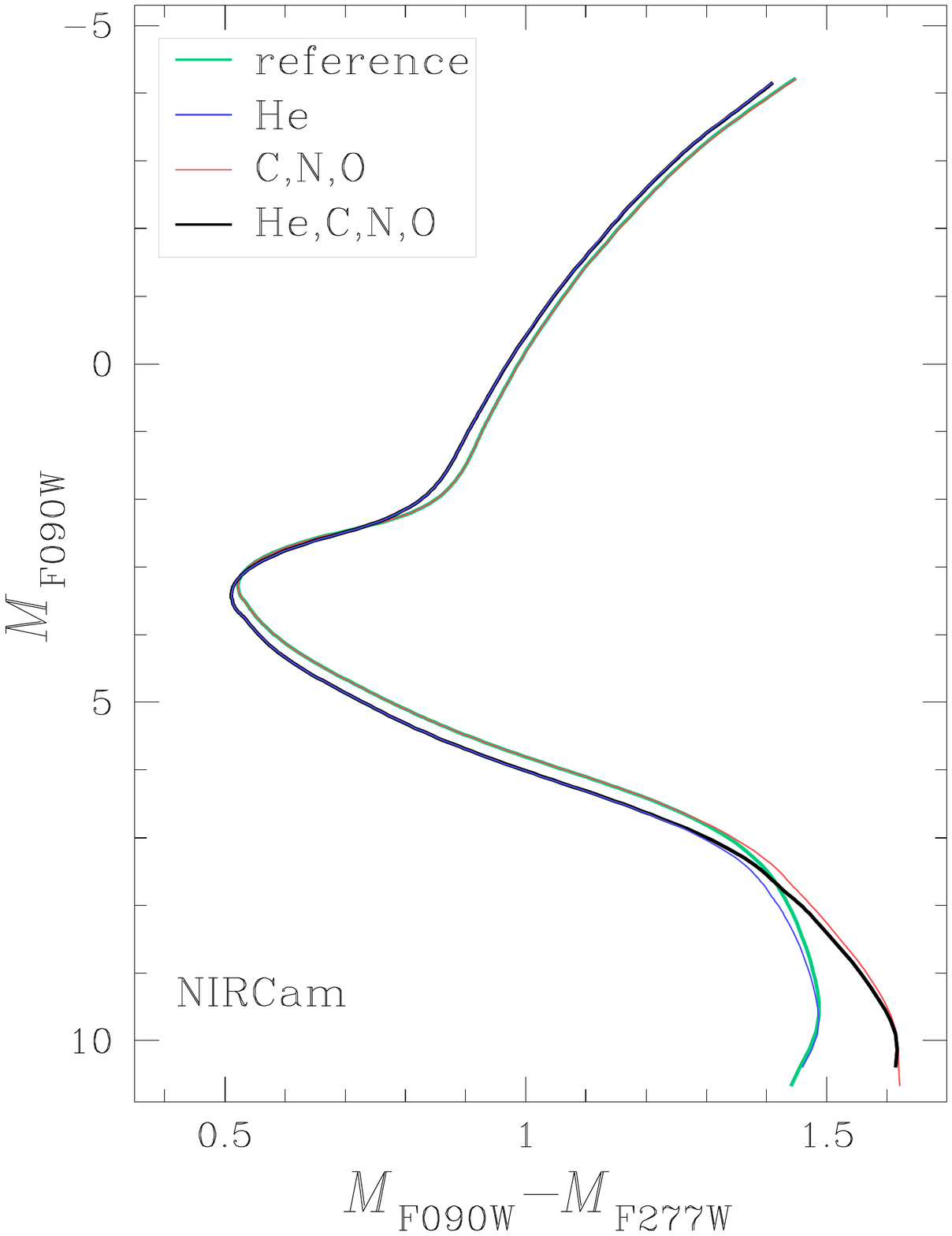} 
    \includegraphics[height=7.6cm,trim={0.6cm 5cm 4.7cm 4.5cm},clip]{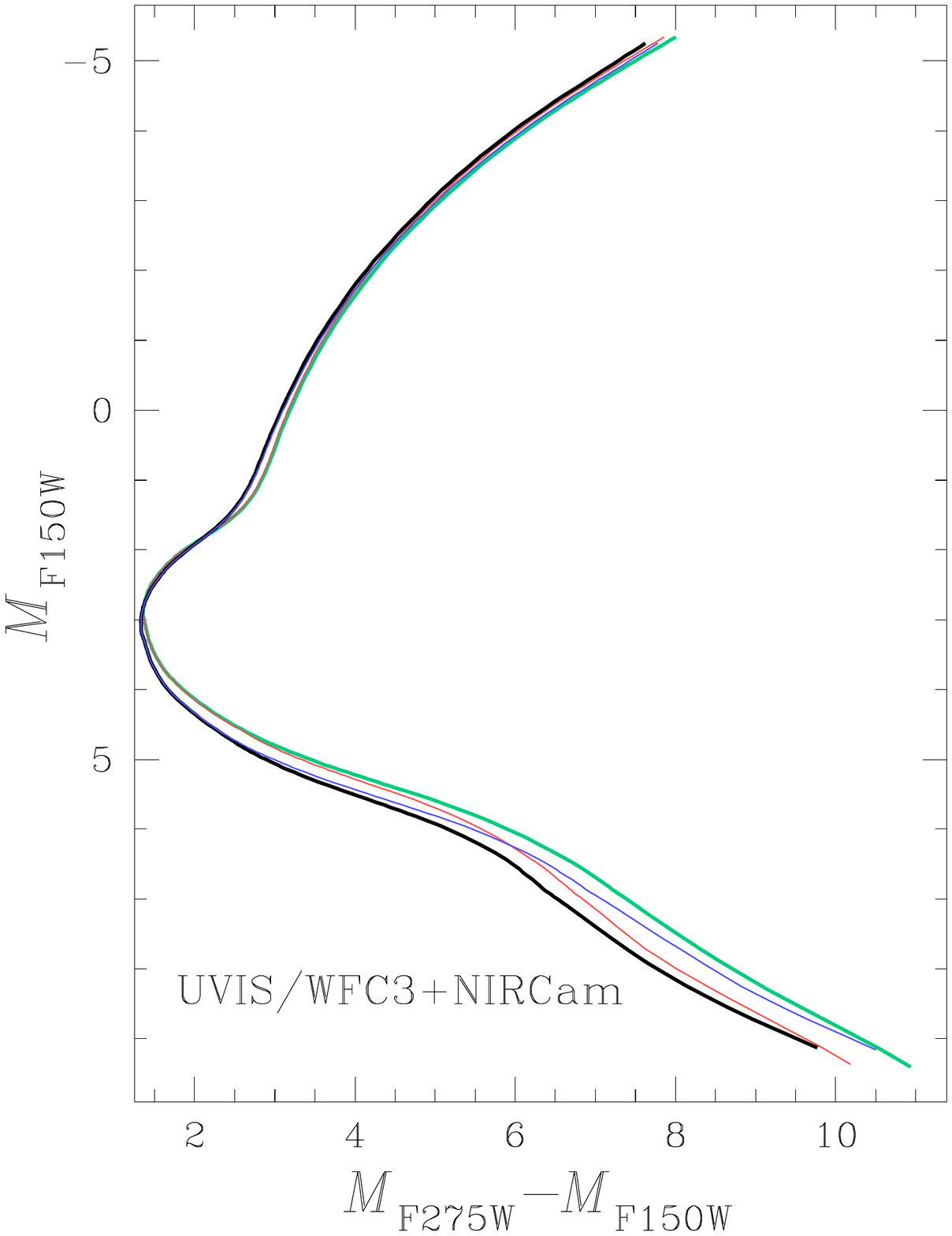} \\
   \includegraphics[height=7.6cm,trim={0.6cm 5cm 4.7cm 4.5cm},clip]{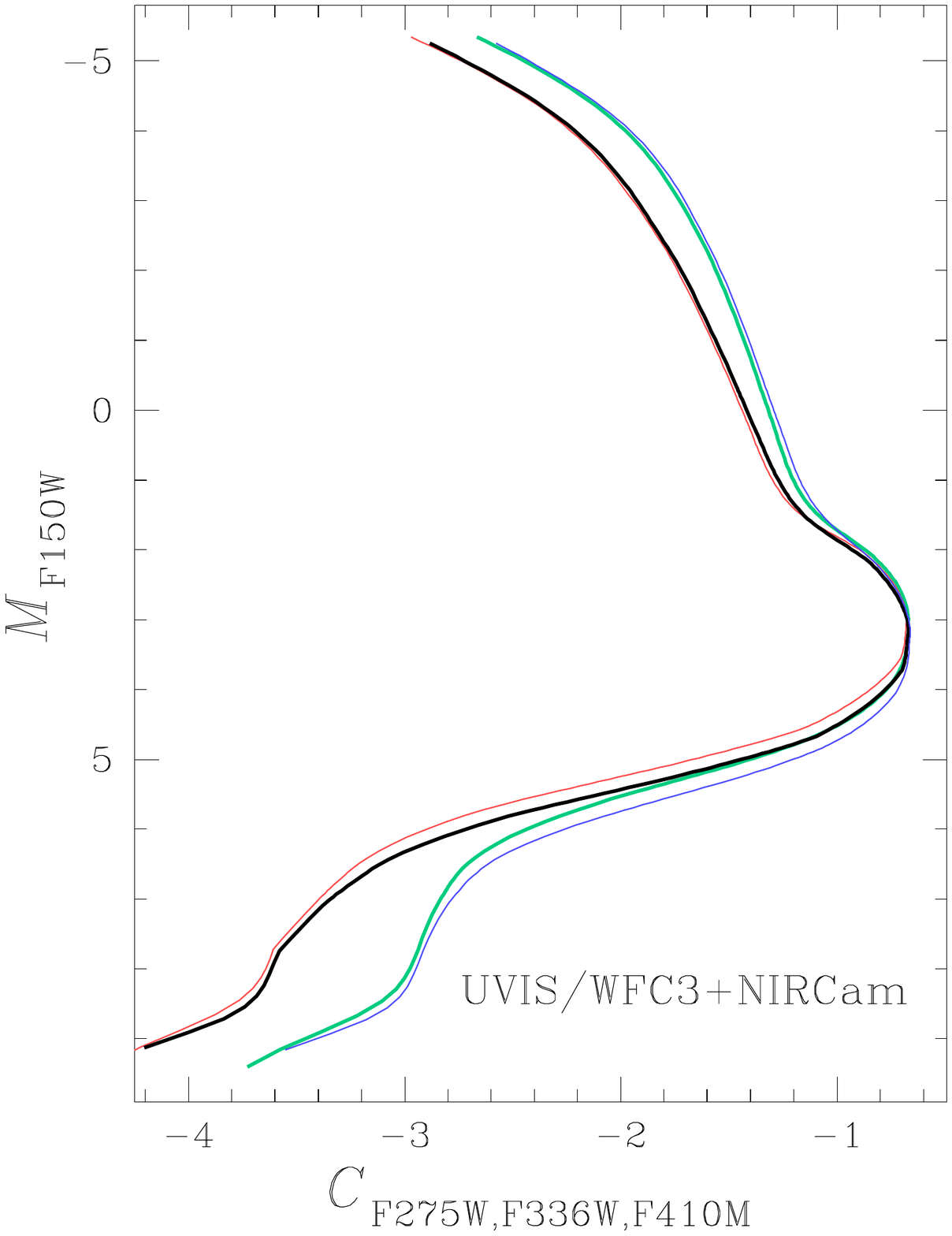} 
    \includegraphics[height=7.6cm,trim={0.6cm 5cm 4.7cm 4.5cm},clip]{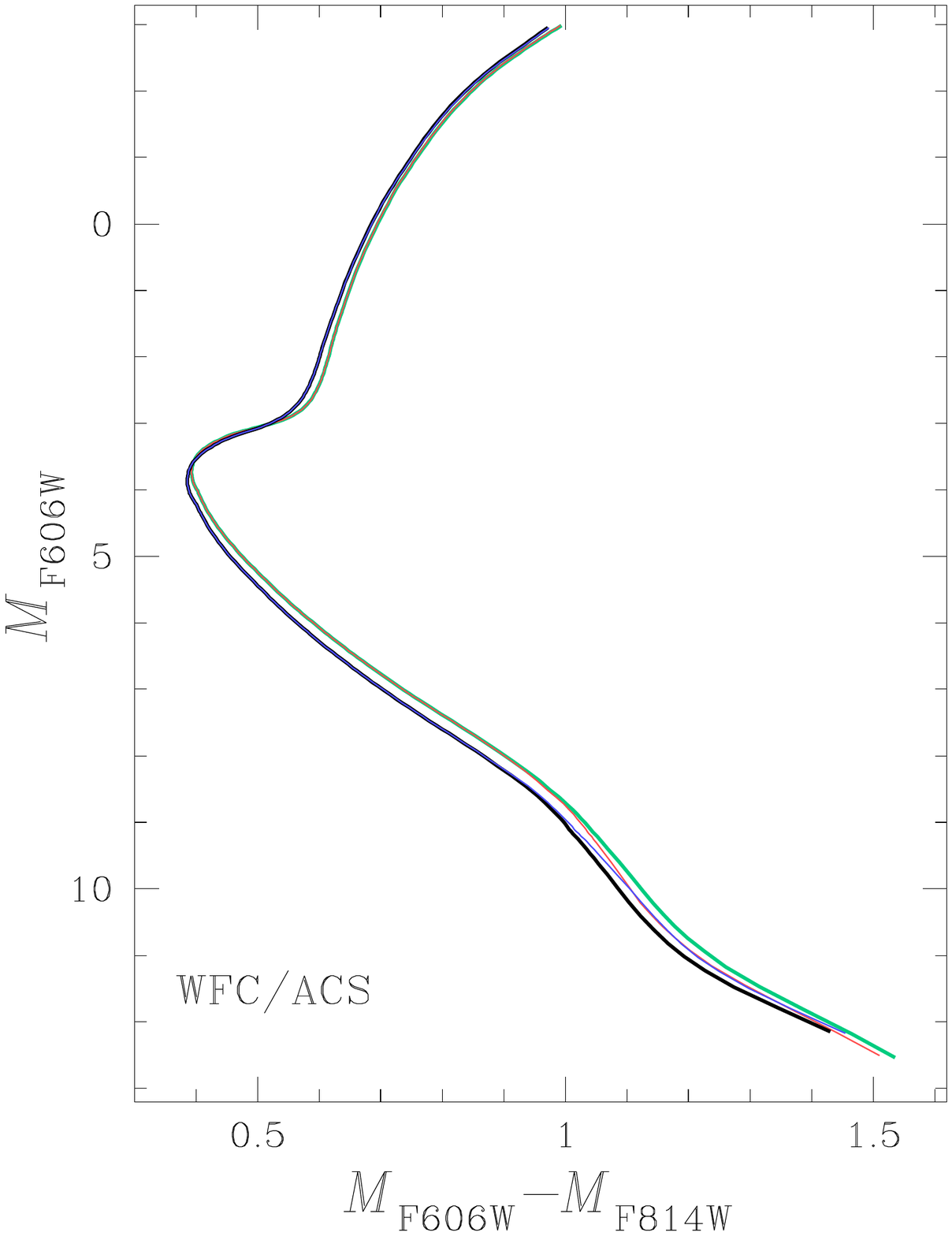} 
  \caption{ Isochrones with ages of 13 Gyr, [Fe/H]=$-$2.3, [$\alpha$/Fe]=0.4, and different abundances of He, C, N, and O. 
  The reference isochrones are colored aqua  and have the same chemical composition of 1G stars. Specifically, they have Y=0.246, [C/Fe]=0.0, [N/Fe]=0.0, and [O/Fe]=0.4. The blue isochrones has the same C, N, O abundances as the reference isochrones but are helium enhanced (Y=0.33). 
  Pink and black isochrones have all [C/Fe]=$-$0.5, [N/Fe]=1.2, and [O/Fe]=$-$0.1 and helium mass fractions of Y=0.246 and 0.33, respectively.}
 \label{fig:iso1} 
\end{figure*} 
\end{centering}

The magnitudes of the 2G stars are derived by adding to the isochrones the corresponding values of $\delta m_{\rm X}$.
 Some results are presented in Figure \ref{fig:iso1}, where we show the isochrones in various photometric diagrams. 
  The pink isochrones represent 1G stars (Y=0.246, C/Fe]=0.0, [N/Fe]=0.0, and [O/Fe]=0.4), whereas black isochrones are helium enhanced with respect to the pink ones (Y=0.33).
   Blue isochrones are indicative of 2G stars with extreme helium content (Y=0.33). They are enhanced in nitrogen by 1.2 dex and depleted in both carbon and oxygen by 0.5 dex with respect to the 1G. Aqua isochrones are similar as the blue ones but have Y=0.246.

The $M_{\rm F090W}$ vs.\,$M_{\rm F090W}-M_{\rm F277W}$ CMD shown on the top-left panel of Figure \ref{fig:iso1} is  sensitive to oxygen variation on M-dwarf atmospheres. Among the CMDs constructed with NIRCam photometry alone, it provides a wide color separation between the photometric sequences of M-dwarfs with different C, N, and O abundances. The reason is that the stellar flux in the F277W band is strongly absorbed by molecules that include oxygen such us H$_2$O and TiO, whereas the F090W filter is poorly affected by these molecules. As a consequence, the 2G stars, which are O-rich exhibit bluer $M_{\rm F090W}-M_{\rm F277W}$ colors that 1G stars with the same luminosity.
 For the same physical reason, the $M_{\rm F090W}-M_{\rm F300M}$ color would provide even wider color separation, although F300M observations would need longer exposure times than the F277W ones to obtain the same signal-to-noise ratio. Similarly, the $m_{\rm F115W}$ vs.\,$M_{\rm F115W}-M_{\rm F322W2}$ CMD would be less sensitive to multiple populations than the $M_{\rm F090W}$ vs.\,$M_{\rm F090W}-M_{\rm F277W}$ CMD but it could be preferable due to the shorter exposure times needed to obtain a given signal-to-noise ratio.

We note that the MSs of stellar populations with different C, N, and O content are nearly superimposed on each other above the MS knee. The isochrones with Y=0.33, which are bluer than the isochrones with primordial helium abundance, provide remarkable exceptions. However, a large difference in helium mass fraction of $\sim 0.076$ corresponds to a $M_{\rm F090W}-M_{\rm F277W}$ color difference of less than 0.01 mag.

\subsection{The helium abundance of multiple populations in M92}\label{subsec:He}
To estimate the abundance variations of He, C, N, and O of 2G$_{\rm A}$ and 2G$_{\rm B}$ stars with respect to the 1G, we followed a procedure widely used in papers from our group \citep[e.g.\,][]{milone2012b, milone2018a, lagioia2019a, zennaro2019a}.

\begin{figure}
    \centering
    \includegraphics[width=8.5cm]{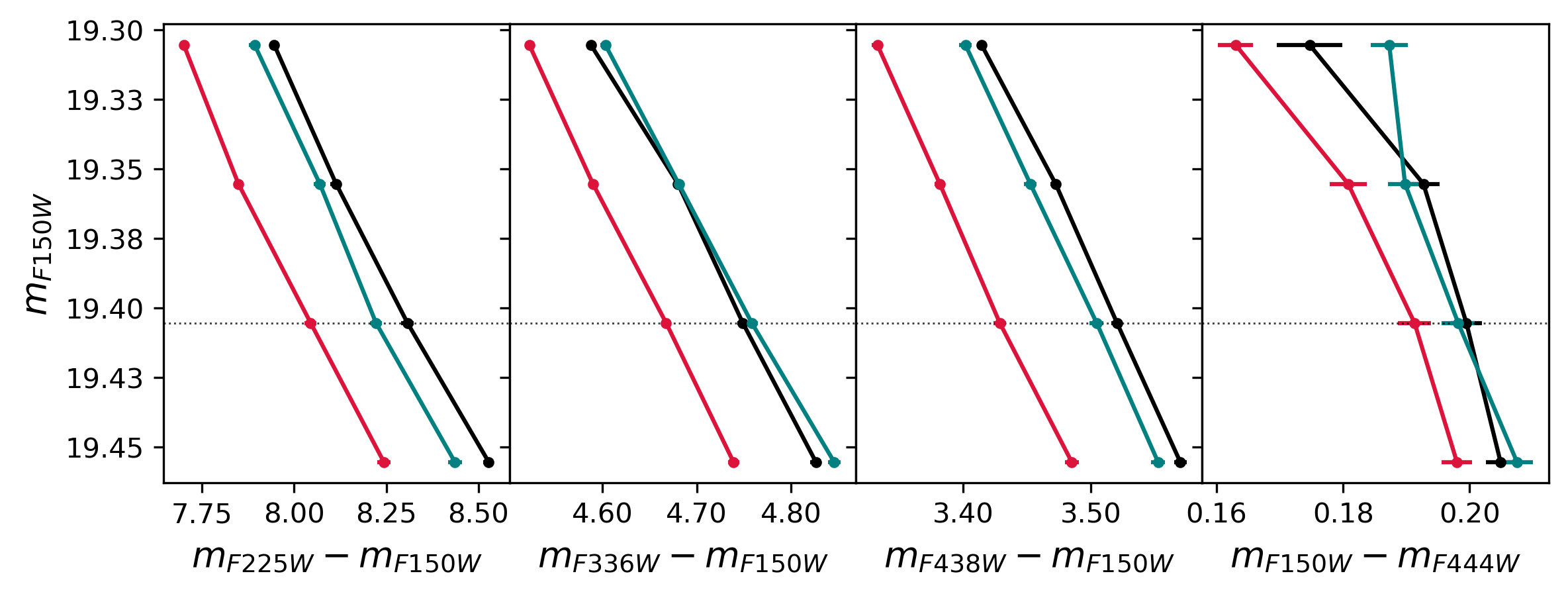}\\
    \includegraphics[width=8.5cm]{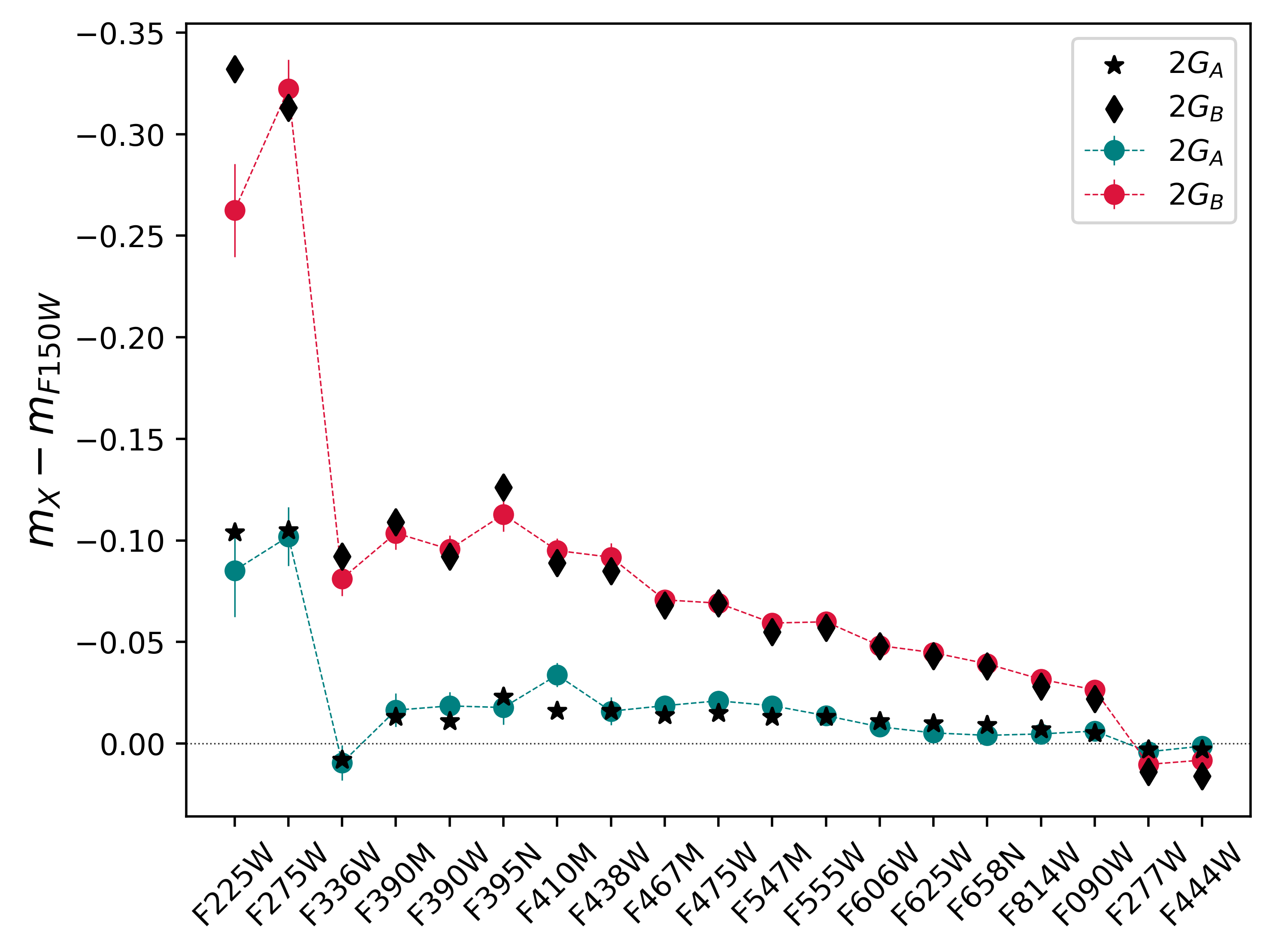}
    \caption{\textit{Top.} Fiducial lines of 1G (black) $2G_{\rm A}$ (green) and $2G_{\rm B}$ (red) stars in the $m_{\rm F150W}$ vs.\,$m_{\rm X}-m_{\rm F150W}$ (or $m_{\rm F150W}-m_{\rm X}$) planes. Here we use X=F225W, F336W, F438W, and F444W. \textit{Bottom.} $m_{\rm X}-m_{\rm F150W}$ color differences relative to the 1G fiducial, for 2G$_{\rm A}$ (green) and 2G$_{\rm B}$ (red) fiducials. The latter corresponds to at $m_{F150W}$=19.41, while the filter names, X, are indicated on the x-axis. Black points represent theoretical predictions for $2G_{\rm A}$ and $2G_{\rm B}$ stars.}
    \label{fig:he}
\end{figure}

We derived the fiducial lines of 1G, 2G$_{\rm A}$, and 2G$_{\rm B}$ stars in the $m_{\rm F150W}$ vs. \,$m_{\rm X}-m_{\rm F150W}$ (or $m_{\rm F150W}$ vs. \,$m_{\rm F150W}-m_{\rm X}$) CMDs, where X indicates the {\it HST} and {\it JSWT} filters listed in Table\,\ref{tab:data}.
 The fiducial lines are calculated in the luminosity interval $18.75<m_{\rm F150W}<19.75$ and are illustrated in the top panels of Figure \ref{fig:he} in the CMDs constructed with the   F225W, F336W, F438W, and F444W filters.
  We identified three equally-spaced magnitude values in the analyzed magnitude interval (dotted horizontal lines in Figure \ref{fig:he}) that we used as reference magnitudes, $m_{\rm ref}$.
   For each value of $m_{\rm ref}$, we calculated the color difference between the fiducials of 2G$_{\rm A}$ and 2G$_{\rm B}$ stars and the 1G fiducial. 
   
   As an example, Figure \ref{fig:he} shows the color differences for the 19 available X filters corresponding to $m_{\rm ref}=$19.41.
   The color separation between bonafide 2G$_{\rm B}$ and 1G stars is maximum for X=F225W and F275W, where it is larger than 0.3 mag, and steadily decreases toward red filters. The F336W, F390M, and F390W filters, which provide narrower color differences than $m_{\rm F395N}-m_{\rm F150W}$, are remarkable exceptions. Noticeably, the F277W and F444W filters provide small color differences between 2G$_{\rm B}$ and 1G stars of about 0.01 mag. The color differences between 2G$_{\rm A}$ stars and 1G stars follow the same qualitative behavior but never exceed $\sim$0.1 mag.
 
We used the best-fitting isochrone to extract the atmospheric parameters, $T_{\rm eff}$ and $g$, corresponding to each value of $m_{\rm ref}$ of 1G and 2G stars.
We computed a grid of synthetic spectra for 2G stars with different abundances of He, C, N, and O, by using the procedure of Section\,\ref{sec:teo}. Each spectrum is compared with the corresponding spectrum of the 1G star, which has Y=0.246, [C/Fe]=0.0, [N/Fe]=0.0, [O/Fe]=0.4.
We assumed for 2G spectra a set of values in [C/Fe]  that range from $-$0.5 to 0.1  dex in steps of 0.05 dex, [N/Fe] between 0.0 and 1.5 in intervals of 0.1 dex, and  [O/Fe] that varies from $-$0.3 to 0.4 in steps of 0.05 dex. The adopted helium abundance varies from Y=0.246 to 0.330 in steps of 0.001 and the values of $T_{\rm eff}$ and $g$ are derived from the Dartmouth isochrone \citep{dotter2008a}.
 The synthetic spectra are convoluted with the transmission curves of the filters used in this paper to derive the corresponding magnitudes. The simulated color differences between the comparison spectra and the reference spectra are compared with the observed color difference between 2G and 1G stars. 
 We assumed that the abundances of He, C, N, and O correspond to the elemental abundances of the comparison synthetic spectrum that best reproduces the observed color differences. Specifically, helium abundances are constrained from the colors constructed with optical and NIR filters, which are mostly sensitive to the effective-temperature difference associated to helium variations. Nitrogen is mainly constrained by the F336W magnitude, which encompasses NH molecular bands, while carbon is derived from the magnitudes in the blue filters, which enclose CH and CN molecules. The F225W and F275W magnitudes are sensitive to oxygen variations through the OH molecules.

We find that 2G$_{\rm A}$ stars are enhanced in helium mass fraction by $\Delta$Y=$0.010 \pm 0.003$ with respect to the 1G. 
 Moreover, 2G$_{\rm A}$ stars have higher abundances of nitrogen ($\Delta$[N/Fe]=0.4$\pm$0.1) and lower content of carbon and oxygen than 1G stars ($\Delta$[C/Fe]=$-$0.15$\pm$0.05,   $\Delta$[O/Fe]=$-$0.25$\pm$0.05).
 2G$_{\rm B}$ stars have a more-extreme chemical composition than the 2G$_{\rm A}$ stars. They are enhanced in helium by $\Delta$Y=$0.041\pm 0.004$ and in nitrogen by ($\Delta$[N/Fe]=0.8$\pm$0.1) with respect to the 1G. When compared with the 1G, 2G$_{\rm B}$ stars are also depleted in carbon and oxygen by $\Delta$[C/Fe]=$-$0.45$\pm$0.10 and  $\Delta$[O/Fe]=$-$0.50$\pm$0.10. Here, the uncertainties are estimated as the root mean scatter of the three elemental abundance determinations corresponding to the three reference magnitudes.

\subsection{Oxygen variations among low mass stars of M92}\label{subsec:O}
\begin{figure}
    \centering
    \includegraphics[height=7.0cm,trim={0.6cm 5cm 0.2cm 8cm},clip]{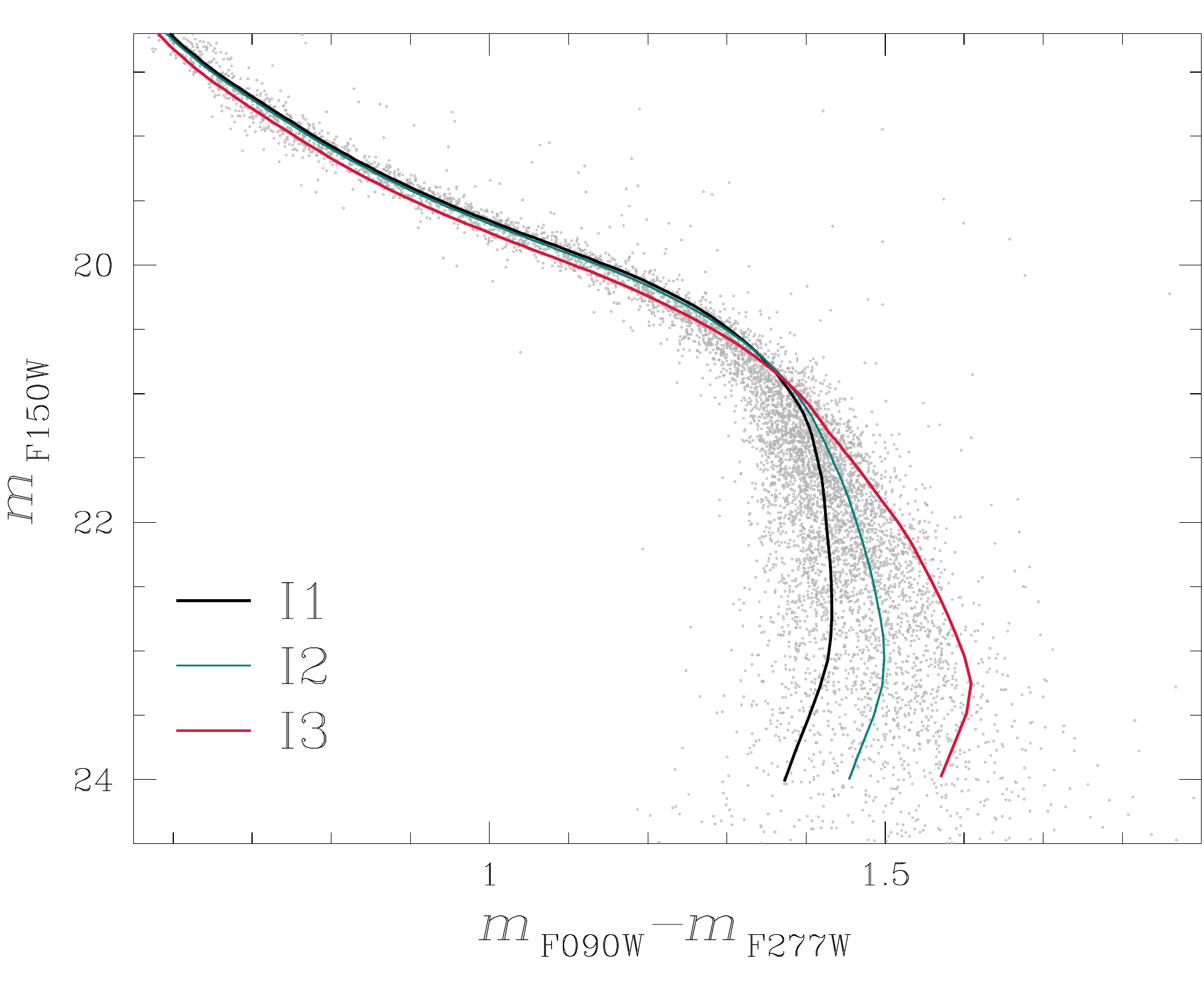}
,    \caption{Best-fitting I1, I2, and I3 isochrones superimposed on the observed $m_{\rm F150W}$ vs.\,$m_{\rm F090W}-m_{\rm F277W}$ CMD. These three isochrones have the same age (13 Gyr) and metallicity ([Fe/H]=$-$2.3) but different contents of He, C, N, and O. Specifically, the chemical composition of I1 isochrone is indicative of 1G stars, whereas I2 and I3 isochrones resemble $2G_{\rm A}$ and $2G_{\rm B}$ stars, respectively. The faintest points in the isochrones correspond to a mass of $0.1 \,M_\odot$.} 
    \label{fig:iso2}
\end{figure}

To constrain the chemical composition of multiple populations among the M-dwarfs, we 
 compared the $m_{\rm F150W}$ vs.\,$m_{\rm F090W}-m_{\rm F277W}$ CMD and 13-Gyr old isochrones with  the same metallicity ([Fe/H]=$-$2.3)  and different light element abundances. Indeed, as discussed in Section\,\ref{subsec:He}, for a fixed luminosity, the  $m_{\rm F090W}-m_{\rm F277W}$ color is very sensitive to oxygen abundance, with O-rich stars showing bluer colors. The results are illustrated in Figure \ref{fig:iso2}, where the black isochrone (I1) represents 1G stars, whereas the green and red isochrones (I2 and I3) have similar content of He, C, N, and O, as  inferred for 2G$_{\rm A}$ and 2G$_{\rm B}$ RGB stars, respectively. 

Although the isochrones do not provide a perfect fit of the MS segment below the knee, the $m_{\rm F090W}-m_{\rm F277W}$ color separation between the I1 and I2 isochrones is similar to the observed MS width, which is consistent with star-to-star oxygen variations of [O/Fe]$\sim$0.5 dex. This fact is consistent with a scenario where stars from the same stellar population (1G or 2G) have the same oxygen abundances along the entire CMD.  Indeed, the oxygen variations in M-dwarfs and bright MS stars derived in this paper are comparable with those inferred for RGB stars by \citet{sneden2000a} and \citet{meszaros2015a}, who obtained a similar oxygen variation of $\sim$0.5 dex by means of high-resolution spectroscopy. 

\section{Summary and discussion}\label{sec:conclusions}
This paper investigates multiple populations in metal-poor GCs by combining {\it HST} and {\it JWST} photometry of M\,92, a metal-poor GC with [Fe/H]=$-2.3$, with isochrones and synthetic spectra.

The photometric study of M\,92 is based on images collected through 16 filters of the UVIS/WFC3, IR/WFC3, and WFC/ACS cameras on board {\it HST} and four NIRCam/JWST filters. The resulting multi-band photometry covers a wide wavelength interval that ranges from the ultraviolet ($\lambda \sim 2,000 \AA$) to the infrared ($\lambda \sim 44,000 \AA$). 
The main results can be summarized as follows.
\begin{itemize}
\item Various photometric diagrams, including the $m_{\rm F150W}$ vs.\,$m_{\rm F275W}-m_{\rm F150W}$ CMD and the $m_{\rm F150W}$ vs.\,$C_{\rm F275W,F336W,F410M}$ pseudo-CMD, show that the color distribution of K-dwarfs along the MS is either intrinsically broad or bimodal, thus revealing multiple populations. 
We combined these two diagrams to introduce the $\Delta_{C \rm F275W,F336W,F410M}$ vs.\,$\Delta_{\rm F275W,F150W}$ ChM, which unveils three main populations of K-dwarf stars: the 1G and two groups of 2G stars, namely 2G$_{\rm A}$ and 2G$_{\rm B}$.
The $m_{\rm F275W}-m_{\rm F150W}$ color separation is mostly associated with differences in helium and oxygen between the stellar populations, whereas oxygen and nitrogen are the main responsibles for the $C_{\rm F275W,F336W,F410M}$ MS width. Indeed, the F275W band encloses OH molecular bands, whereas the F336W filter includes NH bands.

\item We analyzed the internal kinematics of the stellar populations in M\,92 by using stellar proper motions. The three stellar populations  share similar radial distributions of the proper motion dispersion, which range from about 0.2 mas/yr near the cluster center to 0.15 mas/yr around a distance of $\sim$2.2 half-light radii. 1G, 2G$_{\rm A}$, and 2G$_{\rm B}$ stars exhibit isotropic motions in the studied radial interval within $\sim$1.5 half-light radii.
These results corroborate similar conclusions by \citet{libralato2022a, libralato2023a} based on RGB stars.

\item We analyzed the multiple populations in 19 $m_{\rm F150W}$ vs.\,$m_{\rm X}-m_{\rm F150W}$ (or $m_{\rm F150W}-m_{\rm X}$) CMDs, where X indicate the available photometric bands. 
We compared the observed colors of 1G and 2G stars and the colors derived from grids of synthetic spectra to infer the average helium abundance of each population. We find that 2G$_{\rm A}$ and 2G$_{\rm B}$ stars have higher helium mass fractions by $\Delta$Y=0.010$\pm$0.003 and 0.041$\pm$0.004, respectively, than the 1G, for which we assumed a primordial helium abundance of Y=0.246. The helium difference between $2G_{\rm B}$ and 1G MS stars is consistent with the maximum helium variation inferred for RGB stars \citep[$\Delta$Y=0.039$\pm$0.006,][]{milone2018a}.

\item The $m_{\rm F150W}$ vs.\,$m_{\rm F090W}-m_{\rm F277W}$ and the $m_{\rm F150W}$ vs.\,$m_{\rm F090W}-m_{\rm F150W}$ CMDs reveal that MS stars fainter than the MS knee exhibit an intrinsic color spread, which is present among stars with masses of about $\sim$0.1-0.4$\mathcal{M}_{\odot}$. The color broadening is due to stellar populations with different oxygen abundances.
 These low-mass stars exhibit a continuous color distribution and do not show evidence for distinct groups of 1G and 2G stars. 
 The MS width is consistent with star-to-star oxygen variations of [O/Fe]$\sim$0.5 dex. This value is similar to the oxygen difference between 2G$_{\rm B}$ and 1G stars we inferred for K-dwarfs. Moreover, it matches the oxygen interval detected among RGB stars from high-resolution spectroscopy \citep[e.g.][]{sneden2000a, meszaros2015a}. The evidence of multiple populations presenting similar chemical composition among stars with different masses corroborates results obtained for 47 Tucanae \citep{milone2023a}, challenging formation scenarios which predict that GC stars are coeval and the chemical composition of 2G stars is a product of accretion of poluted material onto pre-MS stars \citep[]{gieles2018a}.
\end{itemize}

To investigate the behavior of multiple populations of metal-poor GCs in CMDs constructed with NIRCam filters and with the WFC/ACS and UVIS/WFC3 cameras on board {\it HST}, we derived isochrones with [Fe/H]=$-$2.3 and different abundances of He, C, N, and O. 
Similarly to what we found for metal-intermediate GCs with [Fe/H]=$-$0.75 and [Fe/H]=$-$1.5, the photometric diagrams made with NIRCam filters alone do not allow us to disentangle stellar populations with different abundances of C, N, and O along the RGB, the subgiant branch (SGB), and the MS regions above the MS knee. 2G MS and RGB stars with large helium abundances (Y=0.33) exhibit bluer colors than 1G stars with the same luminosity but their color difference obtained from photometry in NIRCam bands alone is typically smaller than $\sim$0.05 mag.

On the contrary, NIRCam photometry can be a powerful tool to identify 1G and 2G stars below the MS knee, where various colors, including F090W$-$F300M, F090W$-$F277W, and F090W$-$F322W2 allow to identify 1G and 2G M-dwarfs with different oxygen abundances. Optical filters, such as the F070W band of NIRCam or the F606W  filters of WFC/ACS and UVIS/WFC3 on board {\it HST} are poorly sensitive to C,N,O content variations among M-dwarfs. This is in contrast with results of isochrones with [Fe/H]=$-$0.75 and with observations of 47\,Tucanae, which show large F070W and F606W magnitude differences between 1G and 2G stars with different oxygen abundances \citep{milone2023a}.

\begin{acknowledgments}
\section*{acknowledgments} 
This work has received funding from the European Union’s Horizon 2020 research and innovation programme under the Marie Sklodowska-Curie Grant Agreement No. 101034319 and from the European Union – NextGenerationEU, beneficiary: Ziliotto.
 APM and ED  have been supported by MIUR under PRIN program 2017Z2HSMF (PI: Bedin).
 
All of the data presented in this paper were obtained from the Mikulski Archive for Space Telescopes (MAST) at the Space Telescope Science Institute. The specific observations analyzed can be accessed via doi: \dataset[10.17909/7se9-0540]{https://doi.org/10.17909/7se9-0540}.
 
\end{acknowledgments}
\vspace{5mm}

\bibliography{sample631}{}
\bibliographystyle{aasjournal}



\end{document}